\documentclass[journal=jctcce,manuscript=article]{achemso}
\usepackage{graphicx}
\usepackage{rotating}
\usepackage{hyperref}
\usepackage{amsmath}
\usepackage{color}
\usepackage{tabularx}
\usepackage{caption}
\usepackage{epsfig,color}
\usepackage{verbatim}
\usepackage{makeidx}

\usepackage[bb=boondox]{mathalfa}

\SectionNumbersOn

\newcommand{\cm}{cm$^{-1}$}

\def\a0{{$a_{\rm 0}$}}

\newcommand{\p}{^\prime}
\newcommand{\pp}{^{\prime\prime}}

\newcommand{\ket}[1]{\vert #1 \rangle  }
\newcommand{\bra}[1]{\langle #1 \vert  }

\newcommand{\Cv}[1]{${\mathcal C}_{#1{\rm v}}$}
\newcommand{\Dh}[1]{${\mathcal D}_{#1{\rm h}}$}
\newcommand{\Ch}[1]{${\mathcal C}_{#1{\rm h}}$}
\newcommand{\Dd}[1]{${\mathcal D}_{#1{\rm d}}$}

\newcommand{\Cvvv}{${\mathcal C}_{3{\rm v}}$}
\newcommand{\Cs}{${\mathcal C}_{\rm s}$}

\newcommand{\Td}{${\mathcal T}_{\rm d}$}
\newcommand{\Oh}{${\mathcal O}_{\rm h}$}

\newcommand{\2}{$_{2}$}
\newcommand{\3}{$_{3}$}
\newcommand{\4}{$_{4}$}

\newcommand{\schr}{Schr\"{o}dinger}

\newcommand{\Group}{\textit{\textbf{G}}}

\title[TROVE symmetrization approach ]{Symmetry adapted ro-vibrational basis functions for variational nuclear motion calculations: TROVE approach }

\author{Sergei N. Yurchenko}
\affiliation{Department of Physics and Astronomy, University College London, London, WC1E 6BT, UK}
\email{s.yurchenko@ucl.ac.uk}

\author{Andrey Yachmenev}
\affiliation{Center for Free-Electron Laser Science (CFEL), DESY, Notkestrasse 85, 22607 Hamburg, Germany}

\author{ Roman I. Ovsyannikov}
\affiliation{Institute of Applied Physics, Russian Academy of Sciences, Ulyanov Street 46, Nizhny Novgorod, Russia 603950.}

\begin{document}

\date{\today}

\begin{center}
 \leavevmode
\epsfxsize=12.0cm \epsfbox{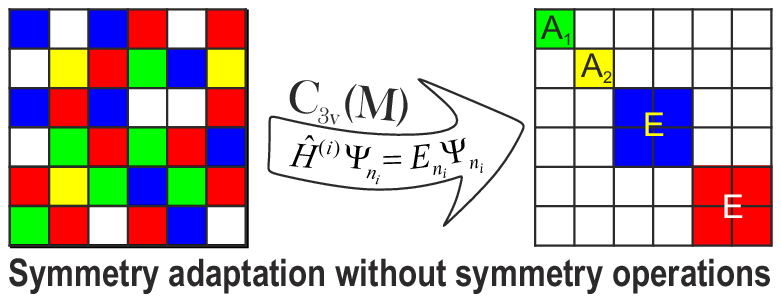}
\end{center}

\begin{abstract}

We present a general, numerically motivated approach to the construction of symmetry adapted basis functions for solving ro-vibrational Schr\"{o}dinger equations. The approach is based on the property of the Hamiltonian operator to commute with the complete set of symmetry operators and hence to reflect the symmetry of the system.
The symmetry adapted ro-vibrational basis set is constructed numerically by solving a set of reduced vibrational eigenvalue problems. In order to assign the irreducible representations associated with these eigenfunctions, their symmetry properties are probed on a grid of molecular geometries with the corresponding symmetry operations. The transformation matrices are re-constructed by solving over-determined systems of linear equations related to the transformation properties of the corresponding wavefunctions on the grid. Our method is implemented in the variational approach TROVE and has been successfully applied to a number of problems covering the most important molecular symmetry groups. Several examples are used to illustrate the procedure, which can be easily applied to different types of coordinates, basis sets, and molecular systems.

%  Diagonal and off-diagonal dipole moment curves are also computed \textit{ab initio}.
\end{abstract}

\maketitle

\section{Introduction}
\label{s:intro}

\begin{figure}[h]
\begin{center}
 \leavevmode
\epsfxsize=8.0cm \epsfbox{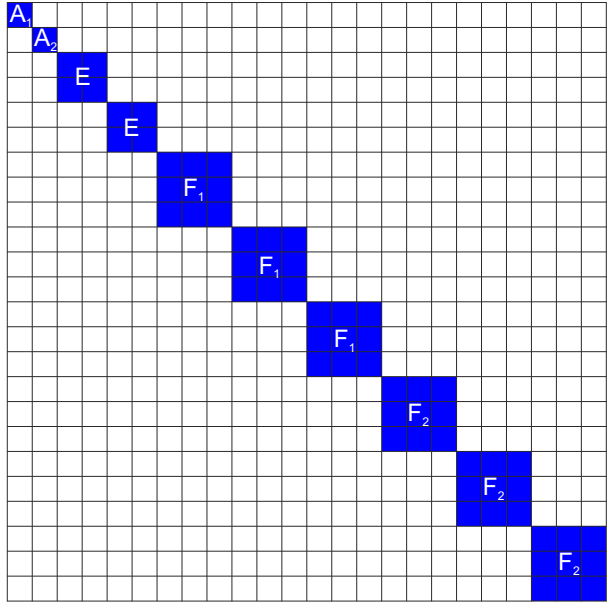}
\caption{\label{f:Td} The block-diagonal structure of a Hamiltonian matrix in the \Td\ irreducible representation. The empty (white) cells indicate vanishing matrix elements. Only 5 blocks of the irreducible representations $A_1$, $A_2$, $E_a$, $F_{1a}$, and $F_{2a}$ are needed as the other five matrices ($E_b$, $F_{1b}$, $F_{1c}$, $F_{2b}$, $F_{2c}$) contain degenerate solutions.}
\end{center}
\end{figure}

Symmetry plays an important role in computing ro-vibrational spectra of polyatomic molecules, particularly in variational solutions of the \schr\ equation. Using a symmetry adapted basis set can considerably reduce the size of the Hamiltonian matrix depending on the symmetry group. For example, in low \Cs\ symmetry (with inversion being the only non-trivial symmetry operation), the use of symmetric and antisymmetric basis functions reduces the matrix by a factor of 2. In higher \Td\ symmetry, the Hamiltonian matrix is split into 10 independent blocks, of which only 5 are needed to determine the unique energies and wavefunctions of the molecular system (see Fig. \ref{f:Td}). For methane, a five-atomic molecule, this is a huge advantage considering the complexity and size of the ro-vibrational computations \cite{13YuTeBa.CH4,14YuTexx.CH4,15NiReTy.method}.

If calculating only the energy levels of a molecule, a symmetry adapted basis set is not essential and any sensible basis should lead to a physically meaningful solution. However, knowledge of the symmetry properties of the eigenvectors is vital for generating spectra, mainly due to the selection rules imposed by the nuclear spin statistics associated with different irreducible representations. Nuclear spin statistical weights give the degeneracy of the ro-vibrational states and contribute to the intensity of a transition. Importantly, some energy levels have zero weights and do not exist in nature. Without knowledge of how the eigenvectors transform under the symmetry operations, it is impossible to describe the molecular spectrum correctly. From a practical perspective, intensity calculations are also much more efficient in a symmetry adapted representation.

The most common symmetry adapted representation is the Wang basis functions, which are simply symmetric and asymmetric combinations of primitive basis functions. Such combinations are sufficient for building symmetrized basis sets for Abelian groups, which consist of one-dimensional irreducible representations only, and this is routinely done in most ro-vibrational applications. It is, however, more challenging to symmetrize the basis set for non-Abelian groups, where the result of the group transformations involve linear combinations of basis functions and cannot be described by simple permutations.
%\Que{Lines 14-18. P3. What do the authors mean when they say that for non-Abelian
%groups the group operators cannot be described by permutation of basis functions?
%Explain why. It is not clear.} \Ans{Done!}
There exist only a handful of ro-vibrational methods in the literature capable of dealing with multidimensional symmetry group representations. Some examples of the variational approaches include works by \citet{03CeSpxx,04BoChGa.method,05YuCaJe.NH3,07YuThJe.method,15PaYuTe.methods,15NiReTy.method,15CaRoIl.CH4,17FaQuCs}.

TROVE (Theoretical ROVibrational Energies)\cite{07YuThJe.method,15YaYuxx.method} is a general method and an associated Fortran 2003 program for computing ro-vibrational spectra and properties of small to medium-size polyatomic molecules of arbitrary structure. It has been applied to a large number of polyatomic species \cite{09YuBaYa,11YaYuJe.H2CO,14SoHeYu.PH3,15SoAlTe.PH3,14UnYuTe.SO3,15AlYuTe.H2CO,14YuTexx.CH4,15YaYuxx.method,15AlOvPo.H2O2,15OwYuYa.CH3Cl,15OwYuYa.SiH4,15AdYaYuJe.CH3,15OwYuTh.NH3,16AlPoOv.HOOH,16UnTeYu,16OwYuYa,16OwYuYa.CH4}, most of which are characterized by a high degree of symmetry (\Cv{3}, \Dh{2}, \Dh{3} and \Td\ symmetry groups). TROVE has proven very efficient for simulating hot spectra of polyatomic molecules and is one of the main tools of the ExoMol project \cite{12TeYuxx.db}. The most recent updates of TROVE have been reported in Ref.~\citenum{16TeYuxx,GAIN}.  Because of the importance of symmetry in intensity calculations, TROVE uses an automatic approach for building the symmetry adapted basis set. In this paper we layout the TROVE symmetrization approach, which is a variation of the matrix symmetrization method.

The matrix symmetrization can be traced back to the original works by \citet{64Gabriel,65MoMoxx,68MoMoxx,67Moccia} and was later extensively developed in a series of papers by, for example, \citet{70DeGuZe,72ChGoxx,72BoGoxx,81JoOrxx,85ChGaMa.method}.
The main idea of these studies is to use a diagonalizion of matrices representing specially constructed symmetry operators.
%specially constructed symmetric matrices.
%These matrices represent operators resembling the symmetry of the system in a suitable basis.
Using this technique a symmetry adaptation can be obtain without the use of symmetry operations \citep{81JoOrxx}.  For example, \citet{67Moccia} used the nuclear attraction matrix to build symmetry adapted molecular orbitals, or a Wilson ${\bf G + F^{-1}}$ matrix in symmetrized force constants calculations; \citet{70DeGuZe}  used  a kinetic ${\bf G}$-matrix to obtain symmetry adapted representations of vibrational molecular modes; \citet{72ChGoxx} used an overlap matrix of atomic orbitals to symmetrize them. \citet{85ChGaMa.method} proposed an `eigenfunction' method based on eigenfunctions of a linear combination of symmetry operations from the so-called complete set of commuting operators (CSCO), which was then extensively employed  for constructing symmetry adapted representations of coordinates and basis functions for (ro-)vibrational calculations\citep{91Katriel,03Lemus.method,11AlLeCa.CH4,12Lexxxx.methods}.

Here we apply the idea of the matrix symmetrization  to numerical construction of  symmetry adapted ro-vibrational representations of a ro-vibrational Hamiltonian  $\hat{H}$ for a general polyatomic molecule. In our version of this method, the symmetry adapted basis functions are generated as eigenvectors of some reduced rovibrational Hamiltonians. %in appropriate non-symmetrized representations.
These operators $\hat{H}^{\rm (red)}$ are derived from $\hat{H}$ such that  (i) they represent different vibrational or rotational modes and (ii) they are symmetrically invariant to $\hat{H}$. According to the matrix symmetrization method, the eigenvectors of $\hat{H}^{\rm (red)}$ necessarily transform according to irreducible representations (irreps) of the symmetry group.

%It makes no sense to use the total ro-vibrational Hamiltonian operator $\hat{H}$  for this %purpose. Instead, we define a set of reduced Hamiltonian operators $\hat{H}^{\rm (red)}$ derived %from $\hat{H}$ by neglecting all vibrational degrees of freedom except ones that form a group in %their own right, which is a subgroup of the symmetry group in question.
%\Que{2.- Lines 17-20. P4. ”form a group in their own right, which is a subgroup of the
%total symmetry group”. This is quite imprecise. The symmetry group is unique. It does
%not exist a ”total” symmetry group. May be the authors wanted to say that dealing
%with the reduced Hamiltonian it is enough to consider a subgroup of the symmetry
%group?} \Ans{rephrased}

Not only does this allow us to construct the symmetry adapted basis functions, but also to improve and contract the basis set via standard diagonalization/truncation procedures. The relative simplicity of this procedure means it can be straightforwardly implemented in many existing nuclear motion programs. It may also be interesting to apply the method in quantum chemical approaches, where the initial set of symmetry adapted atomic orbitals can, for example, be constructed by diagonalizing the bare nuclear Hamiltonian.

%However there are two difficulties in this approach: (i) the symmetry of the eigensolution is not known and needs to be %re-constructed and (ii) the degenerate solutions are arbitrary mixtures of the corresponding components and do not necessarily %transform according with the standard transformation rules \red{WE NEED TO DEFINE THE STANDARD HERE}. In order to resolve these two %issues we sample the symmetry properties of the obtained vibrational or rotational eigen-solutions on a set of grid points, %reconstruct the corresponding transformation matrices $D[R]$ and use the standard symmetry projection technique (see, for example, %\citep{98BuJexx.method}) to reduce the degenerate eigenfunctions to the standard representations. This will also automatically give %the symmetry $\Gamma$ for all eigenfunctions as a by product.

The explanation of our method will be given in the form of practical illustrative examples, rather than using rigorous group-theoretical formalism.
The paper is structured as follows: The main idea of the TROVE symmetrization approach is described in Section~\ref{s:general}. Sections~\ref{s:XY2} and \ref{s:XY3} present illustrative examples for XY\2\ and XY\3\ type molecules.
Readers interested in implementation of the method should read Section~\ref{s:sampling}, where the sampling technique for reconstructing the symmetry transformation properties of vibrational wavefunctions is introduced, and Section~\ref{s:projection}, which details the TROVE reduction method based on the projection operator approach. A non-rigid, ammonia-type molecule XY\3\ of \Dh{3}(M) molecular symmetry is used as an example to illustrate this part of the method implementation. As a very special case, the degenerate mutidimensional isotropic Harmonic oscillator basis functions are considered in Section~\ref{s:Harmonic} with an example shown in the Appendix. Symmetrization of the rotational and total ro-vibrational basis functions is realized using standard reduction techniques and this is discussed in Sections~\ref{s:rotation} and \ref{s:ro-vib}.

%Descriptions of specific types of internal coordinates used in TROVE are collected in the appendixes.

\section{General description of the method}
\label{s:general}

In order to introduce the TROVE symmetrization approach, we consider a general multidimensional ro-vibrational \schr\ equation
\begin{equation}
\label{e:HPsi}
\hat{H} \Psi^{\rm rv}  = E  \Psi^{\rm rv},
\end{equation}
which is to be solved variationally using the ro-vibrational basis set in a product-form:
\begin{equation}
\label{e:basis-set}
  \Phi_{k,\nu}^{J}(\theta,\phi,\chi,q_1,q_2,\ldots,q_{N}) = \ket{J,k,m} \phi_{n_1}(q_1) \phi_{n_2}(q_2) \ldots \phi_{n_N}(q_N),
\end{equation}
where  $\phi_{n_i}(q_i)$ is a one-dimensional (1D) vibrational function,
$n_i$ is a vibrational quantum number, $q_i$ is a generalized vibrational
coordinate, $N$ is the number of vibrational degrees of freedom,  $\ket{J,k,m}$ are the rigid-rotor
wavefunctions,  $k = -J\ldots J$ and $m=-J\ldots J$ are the rotational quantum numbers (projections of the total angular
momentum onto the molecule-fixed $z$ and laboratory-fixed $Z$ axes,
respectively), $\nu= \{n_1,n_2,\ldots n_N\}$ is a generalized vibrational multi-index.
%($\lambda = \{n_1,n_2,\ldots n_N\}$).
The primitive basis functions  $\phi_{n_i}(q_i) \equiv \ket{n_i}$ are any vibrational 1D
functions from a orthonormal set (e.g. Harmonic oscillator wavefunctions). In the absence of external fields $m$ does not play any role and can be omitted.
%\Que{3.- Lines 14-16. P6. What is the meaning of quasi complete orthonormal set?
%Please precise the meaning.} \Ans{rephrased}

%\Que{4.- Line 22. P6. Calling h the order of the group G is not convenient in accordance
%with standard group theoretical language. It is confusing. I suggest to use g (like in the
%classic Hammermesh’s book).}
%\Ans{Corrected here, now to replace h with g everywhere!}
Let us assume that the molecule belongs to a
molecular symmetry\cite{98BuJexx.method} group \Group\ consisting of $g$ elements (group operations) $R$. We aim to construct symmetry adapted basis set functions $\Psi_{\mu}^{J,\Gamma_s}$ which transform according
to irreducible representations $\Gamma_{s}$ ($s=1\ldots r$) of \Group. Here $\mu$ is a counting number and $\Gamma_s$ will be referred to as a `symmetry' or an `irrep' of \Group. For an $l_s$-fold degenerate irrep, and when we will need to refer to specific degenerate components of $\Psi_{\mu}^{J,\Gamma_s}$, an additional subscript $n = 1,\ldots, l_s$ will be used as, e.g.  $\Psi_{\mu,n}^{J,\Gamma_s}$. For example, for the two-fold degenerate $E$ symmetry, $n=1$ and 2 corresponds to the $E_a$ and $E_b$ symmetry components, while in case of the three-fold degenerate $F$ symmetry, these are $F_a$, $F_b$ and $F_c$.
Additionally, we will require that the transformation properties of multi-fold irreps  (e.g. $E$ or $F$ representations) are known.
%\Que{Line 24. P6. The function Psi is supposed to be transformed according to to
%irreps Gammai. But the functions 	Psi themselves do not carry the i index, which by the
%way at this point is not defined. Here Gammai seems to mean i-th irrep. But later an i is
%used for a class. It is confusing. In additions a subindex is lacking concerned with the
%component of the irrep } \Ans{Corrected}

% by a combination of equivalent rotations (SO(3) group):
%\begin{equation}\label{e:D[R]}
%  R\Psi_{m}^{J,\Gamma} = \sum_{n} D[R]_{mn} \Psi_{n}^{J,\Gamma},
%\end{equation}
%where $m$ and $n$ label the degenerate components and  ${\bf D}[R]$ are the standard SO(3) %representation matrices.

We now assume that the symmetry adapted basis functions $\Psi_{\mu,n}^{J,\Gamma_s}$ can be represented by  linear combinations of the sum-of-product primitive functions from Eq.~\eqref{e:basis-set} by
%\Que{6.- Eq.(3). What is the meaning of the index i? A multiplicity index? If this is the
%case a subindex is still lacking corresponding to the component of the irrep. It seems
%to me that the index i is used for several things.}
\begin{equation}\label{e:Psi-vs-Phi}
   \Psi_{\mu,n}^{J,\Gamma_s} = \sum_{k,\nu} T_{k,\nu,n}^{\mu,J,\Gamma_s} \Phi^{J}_{k,\nu} \; ,
\end{equation}
where $T_{k,\nu,n}^{\mu,J,\Gamma_s} $ are  symmetrization
coefficients. The important advantage of the symmetry adapted basis set is that
the corresponding Hamiltonian matrix has a block-diagonal form  (see Fig.~\ref{f:Td}):
\begin{equation}
\label{e:H-block}
\bra{\Psi_{\mu,n}^{J,\Gamma_s}} H^{\rm rv} \ket{\Psi_{\mu',n'}^{J,{\Gamma_t}}} = H_{\mu,\mu'}  \delta_{s,t} \delta_{n,n'} .
\end{equation}
In practice, this means that each  $(J,\Gamma_s,n)$-block can be diagonalized independently with  $J$ and $(\Gamma_s,n)$ as good quantum `numbers' (i.e. constants of motion).
%\Que{Lines 54-56. P6. Good quantum numbers and constant of motions are redundant
%concepts. One implies the other.} \Ans{Corrected}
The main goal of this work is to present a general numerical algorithm  for
constructing symmetrization coefficients $T_{k,\nu,n}^{\mu,J,\Gamma_s} $ for a molecule of general structure and symmetry.

According with the matrix symmetrization method (see, for example, \citet{81JoOrxx}), symmetry adapted set of wavefunctions can be constructed by diagonalizing matrices representing some operators $\hat{A}$. These operators are chosen to be invariant to the symmetry operations $R\in$ \Group.
Our approach is based on the realization that in principle $\hat{H}$ itself would be an ideal choice for $\hat{A}$, as it has the right property to commute with any $R$ from \Group\footnote{Here we assume that there exists isomorphism between the elements $R$ of \Group\ and the corresponding representations, and use the same symbol $R$ in both cases.}
\begin{equation}\label{e:[H,R]}
 [\hat{H},R] = 0.
\end{equation}
%and thus to share the eigenfunctions with any operator $R$.
Indeed, the eigenfunctions of $\hat{H}$ are also eigenfunctions of $R$ (up to a linear combination of degenerate states) and hence transform as one of the irreps of the system (see, for example, the textbook by \citet{89Hamermesh}).
Obviously, it makes no sense to use the ro-vibrational Hamiltonian operator $\hat{H}$  for this purpose. Instead, we define a set of reduced Hamiltonian operators $\hat{H}^{(i)}$ derived from $\hat{H}$ as follows. (i) All ro-vibrational degrees of freedom are divided into $L$ symmetrically independent subspaces, which form subgroups of \Group. (ii) For each $i$th subspace ($i=1\ldots L$) a reduced Hamiltonian operator $\hat{H}^{(i)}$ is constructed by neglecting or integrating over all other degrees of freedom. (iii) The symmetry adapted wavefunctions for each $i$th subspace are obtained by diagonalizing the corresponding $\hat{H}^{(i)}$. (iv) The total basis set is built as a direct product of the subspace bases and then transformed to irreps using standard reduction approaches.

%\Que{9.- Line 13. P7. ”The eigenfunctions of H are also eigenfunctions of R”. This
%is partially true. The sentence is correct only for Abelian groups. By the way, R is
%used as an element of G as well as an operator acting over the functions. There is an
%isomorphism but they are not equal. This must be clarified.} \Ans{The first point is addressed by %adding the comment about the degenerate irreps, the second is also addressed (?)}
%\Que{10.- Line 17. P7. Please precise ”properly selected Hamiltonian operators”.} \Ans{Re-phrased}

%\red{In the following the symbol $R$ is be used both for operators and their representations.}

%\section{Vibrational part}
%\label{s:vib}

%\Que{11.- Line 23. P7. It is inappropriate to use the word ”class” in a text using tools
%of group theory. Class has a precise meaning, different from the definition used in
%the manuscript. Indeed the appropriate concept to be used is ”invariant subspace” or
%”representation space”.} \Ans{class is changed to a subspace}

%Consider a molecule spanning a symmetry group \Group.

Symmetrically independent subspaces of coordinates are selected such that each subspace contains only the coordinates related by symmetry operations
of the group. For example, the vibrational motion of a molecule XY\2\ spanning the molecular symmetry group \Cv{2}(M) can be described by two stretching and one bending mode, which transform independently and can thus be separated into two subspaces. More specifically, the bond lengths $r_1$ (X--Y$_1$) and $r_2$ (X--Y$_2$)  are two stretching vibrational modes connected through symmetry transformations of the group \Cv{2}(M), which form the subspace~$1$, while the interbond angle $\alpha$ (Y$_1$--X--Y$_2$) belongs to the subspace~$2$, with the transformation properties shown in Table~\ref{t:r1-r2-alpha}.

To explore Eq.~\eqref{e:[H,R]} for constructing a symmetry adapted basis, we define and solve a set of eigenvalue problems for reduced Hamiltonian operators $\hat{H}^{(i)}$.  For each subspace $i$ ($i=1\ldots L$) a reduced eigenvalue problem is given by
%\Que{12.- Eq.(6) is confusing. I suggest to use Q(i) for the set of coordinates belonging to
%the i-th subspace. Psii labels the energy, but there is no label associated with degeneracy.}
%\Ans{Changed!: $Q$ to ${\bf Q}$ and check! However, the degeneracy is undefined at this point and %therefore omitted}
\begin{equation}
\label{e:Hphi-reduced}
   \hat{H}^{(i)}({\bf Q}^{(i)}) \Psi_{\lambda_i}^{(i)}({\bf Q}^{(i)}) = E_{\lambda_i} \Psi_{\lambda_i}^{(i)}({\bf Q}^{(i)}),
\end{equation}
where  ${\bf Q}^{(i)}$ is a set of coordinates $\{q_k,q_l,\ldots\}$ from a given subspace $i$, $E_{\lambda_i}$ is an eigenvalue associated with the eigenfunction $\Psi_{\lambda_i}^{(i)}$ and $\lambda_{i}$ counts all the solutions from the subspace~$i$.
The resulting solutions $\Psi_{\lambda_i}^{(i)}$ should transform according with an irrep $\Gamma_s$ of \Group\ and one of its degenerate components $n$ (holds for $l_s>1$).
To indicate the symmetry of the wave function where necessary  the notation $\Psi_{\lambda_i}^{(i),\Gamma_s}$ will be used, or even $\Psi_{\lambda_i,n}^{(i),\Gamma_s}$  to further specify its degenerate components.
%Assuming that $\Psi_{\lambda_i}^{(i)}$ transforms according with the irrep $\Gamma_s$ of \Group\ and one of its degenerate components $n$ (for $l_s>1$), indices   $\Gamma_s$ and $n$ will be also used to indicate its symmetry as $\Psi_{\lambda_i,n}^{(i),\Gamma_s}$.
% is also given in order to indicate that the eigenfunction is expected to span one of the irreps %of \Group, $\Gamma^{(i)}$, i.e. transform according to $\Gamma^{(i)}$, or one of its degenerate %components for $d_t>1$.
%In fact, $\Psi_{\lambda_i}$ should also include an index $n (n\le d)$, indicating the %corresponding degenerate component, if present, which we omit from Eq.~\eqref{e:Hphi-reduced} for %simplicity.

The reduced Hamiltonian $\hat{H}^{(i)}$ is constructed by  averaging the total vibrational ($J=0$) Hamiltonian $\hat{H}$ on the `ground state' primitive vibrational basis functions $\phi_{n_s}(q_{s}) =  \ket{n_s}$ from other subspaces ($\{s\}\not \in \{i\}$) as given by
 \begin{equation}\label{e:H-reduced}
   \hat{H}^{(i)}({\bf Q}^{(i)}) = \bra{0_p} \bra{0_q} \ldots \bra{0_{r}} \hat{H} \ket{0_r} \ldots \ket{0_q} \ket{0_p},
 \end{equation}
where $\ket{0_s} $  is a primitive basis function $\phi_{n_s}(q_s)$ with $n_s = 0$ and  $\{p,q,r\}$ are coordinates from other subspaces, i.e. $\{p,q,r\} \not \in \{i\}$.

For example, in the case of an XY$_2$ molecule  the two reduced Hamiltonian operators can be formed as
\begin{eqnarray}
\label{e:HXY2:1}
  \hat{H}^{(1)}(r_1,r_2) &=& \bra{0_3} \hat{H} \ket{0_3}, \\
\label{e:HXY2:2}
  \hat{H}^{(2)}(\alpha) &=& \bra{0_1}\bra{0_2} \hat{H} \ket{0_2}\ket{0_1},
\end{eqnarray}
where ${\bf Q}^{(1)}$ = $\{r_1,r_2\}$ and ${\bf Q}^{(2)}=\{\alpha\}$ define the partitioning of the three coordinates into two subspaces $i=1$ and $2$.

%\Que{13.- Line 35. P8. Again the function Psi
% does not present degeneracy label.}  \Ans{Corrected}

Equation~\eqref{e:Hphi-reduced} represents the main idea of the method, which will be referred to as TROVE symmetrization:
since $\hat{H}^{(i)}$ commutes with any $R \in$ \Group,
the eigenfunctions $\Psi_{\lambda_i}^{(i)}({\bf Q}^{(i)})$
must necessarily span one of the irreducible representations $\Gamma_s$ of the group \Group.
%Using eigenfunctions of some specially selected operators from a complete set of commuting %operators for symmetrization has been proposed by \citet{89Chen} and then developed in a series %of works by \citeauthor{12Lexxxx.methods} for ro-vibrational %calculations\citep{03Lemus.method,11AlLeCa.CH4,12Lexxxx.methods}. In their approach, the %eigenvalue problem is solved for an operator  $\hat{C}$ constructed as a linear combination of %symmetry operations $R$ from the so-called complete set of commuting operators (CSCO). Our %eigenfunction approach is based on the Hamiltonian operator $\hat{H}$ itself, which does not %directly belong to CSCO, but commutes with all its members.
By solving Eq.~\eqref{e:Hphi-reduced}, not only do we get a more compact basis set representation which can be efficiently contracted following the diagonalization/truncation approach, it is also automatically symmetrized.
%As we will show below, this method can be applied both to the vibrational and rotational degrees of freedom.
The total vibrational basis set is then constructed as a direct product of $L$ symmetrically adapted basis sets followed by a reduction to irreducible representations using standard projection operator techniques (see, for example, Ref.~\citenum{98BuJexx.method}). The major advantage of this symmetrization approach is that it can be formulated as a purely numerical procedure, which is particularly valuable for handling the algebra of symmetry transformations to describe high vibrational excitations. The required components (Hamiltonian matrices and eigensolvers) are readily available in any variational program and thus the implementation of the present approach into a variational ro-vibrational calculation should be relatively straightforward.

%\Que{14.- Line 12. P9. Please eliminate the adjective ”thorny”. It is a relative adjective
%strongly dependent of the user’s skill on group theory.} \Ans{Done}

There are however two major problems to overcome: (i) Even though we know that $\Psi_{\lambda_i}^{(i)}$ from Eq.~(\ref{e:Hphi-reduced}) should transform as an irrep $\Gamma_s$,  we do not automatically know which one, except for trivial one-dimensional subspaces; (ii) the degenerate solutions (e.g. for $\Gamma_s = E, F, G$) are usually represented by arbitrary mixtures of the degenerate components and do not necessarily transform according to standard irreducible transformation matrices (see also examples below). The latter is a common problem of degenerate solutions, since any linear combination of degenerate eigenfunctions is also an eigenfunction. For example, when TROVE solves Eq.~\eqref{e:Hphi-reduced} by a direct diagonalization using one of the numerical linear algebra libraries (e.g. DSYEV from LAPACK), the degenerate eigenfunctions come out as unspecified mixtures of degenerate components. We will demonstrate in Section~\ref{s:projection} that a general reduction scheme can be used to recast the degenerate mixtures such that they follow the standard transformation properties upon the group operations. It should be noted that the eigenvalue symmetrization method by \citet{03Lemus.method,12Lexxxx.methods} can in principle be used to resolve the degenerate components by constructing a proper CSCO.
%\Que{15.- Line 32-41. P 9. As far as I know, Chen’s approach deals precisely with ob-
%taining symmetry adapted functions through the diagonalization of a CSCO associated
%with a chains of groups. This approach provides functions carrying irreps and compo-
%nents simultaneously with a precise matrix representation. The representation matrix
%is implicit in the calculation of the matrix elements h  0 (R) for the
%generators of the group, a not difficult task in this context. However, in the context of
%Chen’s approach, it is not necessary to construct such matrices. Hence, is questionable
%that the Chen’s method ”does not have this issue”.}

%\subsection{Symmetries via numerical sampling of wavefunctions}

\begin{table}
\begin{center}
\caption{\label{t:r1-r2-alpha} Transformation properties of the internal coordinates $r_1$, $r_2$, and $\alpha$ of an XY$_2$-type molecule and the characters of the irreps of the \Cv{2}(M) group.}
\begin{tabular}{ccccc}
\hline
Coordinate & $E$ & $(12)$ & $E^{*}$ & $(12)^{*}$ \\
\hline
$r_1        $&$ r_1 $&$ r_2 $&$ r_1 $&$ r_2  $\\
$r_2        $&$ r_2 $&$ r_1 $&$ r_2 $&$ r_1  $\\
$\alpha     $&$ \alpha $&$ \alpha $&$ \alpha $&$ \alpha $\\
\hline
Irrep $\Gamma$ & \multicolumn{4}{c}{Characters $\chi$ }\\
\hline
$A_1$        &  1  &  1 &  1 & 1 \\
$A_2$        &  1  &  1 & -1 &-1 \\
$B_1$        &  1  & -1 & -1 & 1 \\
$B_2$        &  1  & -1 &  1 &-1 \\
\hline
\hline
\end{tabular}
\end{center}
\end{table}

\section{Examples}

\subsection{Vibrational basis set for XY\2-type molecules}
\label{s:XY2}

%\Que{2. The numerical examples given are very nice but they require the use of the complex com-
%puter code TROVE. It might be useful to give a few simpler toy examples which the reader
%herself(himself) could check easily by writing her own code.} \Ans{reformulate such that it would %sound general}

In order to demonstrate how TROVE symmetrization based on Eq.~(\ref{e:Hphi-reduced}) works, we again consider an XY\2\ triatomic molecule. It spans the Abelian group \Cv{2}(M) with well-known symmetry adapted combinations of vibrational basis functions given by (compare to Eq.~\eqref{e:Psi-vs-Phi}):
%\Que{16.- Eq.(10-11).
%The labeling in this equations is different from Eq.(2). These func-
%tions are one quantum states or for any quanta? What is the value of ni in accordance
%with Eq.(2)? What is the meaning of i here? If i refers to an invariant subspace, I
%would expect to define i = 1 for stretches and i = 2 for bends or vice versa, but this is
%not reflected in Eqs.(10-11).}
\begin{eqnarray}
\label{e:Phi:XY2:1}
\Phi_{n_1,n_2,n_3}^{A_1} &=& \frac{1}{\sqrt{2}} \left[ \phi_{n_1}(r_1)\phi_{n_2}(r_2) + \phi_{n_2}(r_1) \phi_{n_1}(r_2)   \right]\phi_{n_3}(\alpha), \quad n_1 \ne n_2 \\
\label{e:Phi:XY2:2}
\Phi_{n_1,n_2,n_3}^{B_2} &=& \frac{1}{\sqrt{2}} \left[ \phi_{n_1}(r_1)\phi_{n_2}(r_2) - \phi_{n_2}(r_1) \phi_{n_1}(r_2)
\right]\phi_{n_3}(\alpha), \quad n_1 \ne n_2 \\
\Phi_{n,n,n_3}^{A_1} &=&  \phi_{n}(r_1)\phi_{n}(r_2)\phi_{n_3}(\alpha) , \quad n_1 = n_2 \equiv n,
\end{eqnarray}
where $A_1$ and $B_2$ are two irreducible representations of \Cv{2}(M) (see Table~\ref{t:r1-r2-alpha}).
The `irreducible' functions $\Phi_{n_1,n_2,n_3}^{A_1}$ and $\Phi_{n_1,n_2,n_3 }^{B_2}$ are also eigenfunctions of the group operators $R$  = $\{E,(12),E^{*},(12)^{*}\}$, e.g.
\begin{eqnarray}
(12) \, \Phi_{\nu}^{A_1} &=&  \Phi_{\nu}^{A_1}, \\
(12) \, \Phi_{\nu}^{B_2} &=& -\Phi_{\nu}^{B_2},
\end{eqnarray}
where $\nu$ stands for $\{n_1,n_2,n_3 \}$.
The transformation of  the `reducible' primitive functions $\ket{n_1}\ket{n_2}\ket{n_3}$ = $\phi_{n_1}(r_1)\phi_{n_2}(r_2)\phi_{n_3}(\alpha) $ (for $n_1\ne n_2$), that  are not eigenfunctions of $R=(12)$, involves two different states:
\begin{eqnarray}
(12) \ket{n_1}\ket{n_2}\ket{n_3} &=&  \ket{n_2}\ket{n_1}\ket{n_3}, \\
(12) \ket{n_2}\ket{n_1}\ket{n_3} &=&  \ket{n_1}\ket{n_2}\ket{n_3}.
\end{eqnarray}

%poten_h2s_dvr3d from h2s_dvr3d_sym_01.inp
Now we derive irreducible combinations of $\ket{n_1}\ket{n_2}\ket{n_3}$  using the numerical approach of Eq.~(\ref{e:Hphi-reduced}). As an example here we use the vibrational wavefunctions of the H\2S molecule obtained variationally with TROVE based on the potential energy surface from Ref.~\citenum{16AzTeYu}. It should be noted however that any computational approach using the same coordinates would essentially give equivalent expansions.
%We
%%use TROVE \cite{07YuThJe.method,15YaYuxx.method} and
%build two reduced Hamiltonians (see Eqs.~(\ref{e:HXY2:1},\ref{e:HXY2:2})) in terms of the coordinates $q_1 = \Delta r_1$, $q_2 = \Delta r_2$, $q_3 = \Delta \alpha$
We construct the matrix representations of the reduced Hamiltonians in Eqs.~(\ref{e:HXY2:1}) and (\ref{e:HXY2:2}) in the basis of 1D functions $\phi_{n_1}(r_1)$, $\phi_{n_2}(r_2)$, and $\phi_{n_3}(\alpha)$ determined using the Numerov-Cooley \cite{23Nuxxxx.method,61Coxxxx.method} approach as described in Ref.~\citenum{07YuThJe.method}.
For simplicity we use a small basis set limited by the polyad number $P_{\rm max} = 2$ as given by
$$
P = n_1 + n_2  + n_3 \le P_{\rm max}.
$$

After solving the reduced eigenvalue problem for $\hat{H}^{(1)}$  (Eq.~\eqref{e:HXY2:1}),  the following variational wavefunctions were obtained:
\begin{eqnarray}
\label{e:phi-A1}
 \Psi_{1}^{(1)}(r_1,r_2)  &=  &0.999999 \ket{0,0} +0.0000548 \left( \ket{0,1} + \ket{1,0} \right)
 %-0.003072 \left( \ket{0,2} + \ket{2,0}\right)
 + \ldots
 ,  \\
\label{e:phi-A1-2}
 \Psi_{2}^{(1)}(r_1,r_2)  &=  &0.0000775 \ket{0,0} -0.7071066 \left( \ket{0,1} + \ket{1,0} \right) %-0.003783    \left( \ket{0,2} + \ket{2,0} \right)
 + \ldots
 , \\
\label{e:phi-B2}
 \Psi_{3}^{(1)}(r_1,r_2)  &= &-0.7071068 \left( \ket{0,1} - \ket{1,0} \right)
 + \ldots
 %-0.003783    \left( \ket{0,2} - \ket{2,0} \right)
 ,
\end{eqnarray}
%\begin{eqnarray}
% \Psi_{1}^{A_1}(r_1,r_2)  =  &&0.99999 \ket{0,0} +0.00055
% \left( \ket{0,1} + \ket{1,0} \right)
%+ 0.00005
%\left( \ket{0,2} + \ket{2,0}\right) + 0.000958 \ket{1,1}
% ,  \\
%\label{e:phi-A1-2}
% \Psi_{2}^{A_1}(r_1,r_2)  =  &&0.00078
% \ket{0,0} -0.70711
%\left( \ket{0,1} + \ket{1,0} \right)+ 0.00030
%    \left( \ket{0,2} + \ket{2,0} \right) -0.002230
%\ket{1,1}
% , \\
%\label{e:phi-B2}
% \Psi_{3}^{B_2}(r_1,r_2)  = &&0.70711
% \left( \ket{0,1} - \ket{1,0} \right)
% -0.00085
%  \left( \ket{0,2} - \ket{2,0} \right)
% ,
%\end{eqnarray}
where we have used a shorthand notation $|n_1,n_2\rangle=|n_1\rangle|n_2\rangle$.
When compared to Eqs.~(\ref{e:Phi:XY2:1},\ref{e:Phi:XY2:2}), the eigenfunctions $\Psi_{\lambda_1}^{(1)}$ have the expected symmetrized form and are classified according to the $A_1$ and $B_2$ irreps, i.e. as $\Psi_{1}^{(1),A_1}$, $\Psi_{2}^{(1),A_1}$, and $\Psi_{3}^{(1),B_2}$. Thus the expansion coefficients
$T_{k,\nu,n}^{\mu,J,\Gamma_s} = T_{\{n_1,n_2\}}^{\mu,\Gamma_s} $ in Eq.~(\ref{e:Psi-vs-Phi}) are obtained numerically without any assumption on the symmetries. Here $J,k=0$ (rotational indices) and $n=1$ (degenerate component)  are omitted for simplicity and $\nu=\{n_1,n_2\}$.
The  numerical error of the symmetrization can be assessed by comparing Eqs.~(\ref{e:phi-A1}--\ref{e:phi-B2}) to Eqs.~(\ref{e:Phi:XY2:1},\ref{e:Phi:XY2:2}).  For example, the differences between $T_{\{1,0\}}^{1,A_1}$ and $T_{\{0,1\}}^{1,A_1}$, $T_{\{1,0\}}^{2,A_1}$ and $T_{\{0,1\}}^{2,A_1}$, $T_{\{1,0\}}^{3,B_2}$ and $-T_{\{0,1\}}^{3,B_2}$ are found to be within $10^{-15}$.

Increasing the size of the basis set (using larger $P_{\rm max}$) will lead to analogous expansions involving symmetrized contributions from higher excitations  $\ket{n_1,n_2}$. The new reduced wavefunctions $\Psi_{\lambda_1}^{(1),\Gamma_s}(r_1,r_2)$ together with $\Psi_{\lambda_2}^{(2),\Gamma_s'}(\alpha)$ (eigenfunctions of $H^{(2)}$ in Eq.~(\ref{e:HXY2:2})) can be utilized to build the new contracted and symmetrized basis set, which is then used to diagonalize the complete Hamiltonian $\hat{H}$. In this simple example the symmetry properties of the expansion coefficients, as well as of the corresponding wavefunctions, are trivial. However, our goal is to develop a general \textit{numerical} symmetrization algorithm applicable to arbitrary basis sets, coordinates, symmetries or molecules, which is also in line with the TROVE ideology of a general, black-box like program. As will be demonstrated below, the advantage of our automatic symmetry classification method becomes more pronounced for larger molecules with more complicated symmetry, especially for ones containing degenerate representations.

\subsection{Tetratomics of the XY$_3$-type, \Cv{3}-symmetry}
\label{s:XY3}

Here we present another example of a rigid pyramidal tetratomic molecule XY\3, characterized by the \Cvvv(M) molecular symmetry group.
%Here we use a rigid pyramidal tetratomic molecule XY\3, characterized by the \Cvvv(M) molecular symmetry group, to construct a symmetrized basis set using the TROVE symmetrization approach.
We choose six internal coordinates as $\Delta r_1, \Delta r_2, \Delta r_3$ (bond length displacements) and $\Delta \alpha_{12}, \Delta \alpha_{13}$,  $\Delta \alpha_{23}$ (the interbond angle displacements).
%the generalized internal coordinates $q_i$ as the geometrically defined coordinates $\Delta r_1, \Delta r_2, \Delta r_3$ (bond length displacements) and $\Delta \alpha_{12}, \Delta \alpha_{13}$,  $\Delta \alpha_{23}$ (the interbond angle displacements).
The associated permutation symmetry operations and characters of  \Cv{3}(M) are collected in Table~\ref{t:C3v:r1r2r3}. These coordinates, as well as the corresponding 1D primitive basis functions $\ket{n_i}$ ($i=1\ldots 6$), form two subspaces that transform independently: subspace 1 is  $\{\Delta  r_1, \Delta r_2, \Delta r_3\}$ and subspace 2 is  $\{\Delta \alpha_{12}, \Delta \alpha_{13}, \Delta \alpha_{23}\}$. We assume that $\ket{n_1}$, $\ket{n_2}$, and $\ket{n_3}$ are the 1D stretching basis functions of $\Delta r_1, \Delta r_2$, and $\Delta r_3$, respectively,  and $\ket{n_4}$, $\ket{n_5}$, and $\ket{n_6}$ are the 1D bending functions of $\Delta \alpha_{12}, \Delta \alpha_{13}$, and $\Delta \alpha_{23}$, respectively.  The two reduced \schr\ equations for subspaces 1 and 2 are given by:
%\begin{eqnarray}
%\Que{17.- Eq.(20). Again, it seems to me that degeneracy is systematically omitted,
%unless impossible to exclude, like in Eq.(22).} \Ans{This equation is corrected, but degeneracy %is still omitted since it is still undefined at this point}
\begin{equation}
\label{e:XY3-red}
\begin{split}
  \hat{H}_{\rm str}^{(1)} \Psi_{\lambda_1}^{(1)} &= E_{\lambda_1} \Psi_{\lambda_1}^{(1)}, \\
%\label{e:XY3-red-2}
  \hat{H}_{\rm bnd}^{(2)} \Psi_{\lambda_2}^{(2)} &= E_{\lambda_2} \Psi_{\lambda_2}^{(2)},
%\end{eqnarray}
\end{split}
\end{equation}
where the reduced 3D Hamiltonian operators $\hat{H}_{\rm str}^{(1)}$ and $\hat{H}_{\rm bnd}^{(2)}$ are obtained by vibrationally averaging the total vibrational Hamiltonian $\hat{H}^{\rm 6D}$ over the ground state basis functions from subspace 2 and 1, respectively:
\begin{equation}
\begin{split}
  \hat{H}_{\rm str}^{(1)}(\Delta r_1,\Delta r_2,\Delta r_3) &= \bra{0_4} \bra{0_5} \bra{0_6} \hat{H}^{\rm 6D} \ket{0_6} \ket{0_5} \ket{0_4},  \\
\hat{H}_{\rm bnd}^{(2)}(\Delta\alpha_{12},\Delta\alpha_{13},\Delta\alpha_{23}) &= \bra{0_1} \bra{0_2} \bra{0_3} \hat{H}^{\rm 6D} \ket{0_3} \ket{0_2} \ket{0_1}.
\end{split}
\end{equation}
The \Cv{3}(M) group spans the $\Gamma_s=A_1$, $A_2$, and $E$ representations, where $E$ is two-fold.
Following the discussion above, we expect the resulting wavefunctions $\Psi_{\lambda_i}^{(i)}$ to be eigenfunctions of all six symmetry operators $R$ of \Cv{3}(M) from Table~\ref{t:C3v:r1r2r3}, i.e. to transform according to $A_1$, $A_2$, or $E$.
%Once again, solving the reduced eigenvalue problems in Eqs.~(\ref{e:XY3-red}) will produce symmetrized wave functions.

\begin{table}
\caption{The character table and transformation of the internal coordinates of an XY$_3$-type molecule by the symmetry operations
  of the molecular symmetry group \Cv{3}(M) \cite{98BuJexx.method}.}
%\squeezetable
\small
\begin{center}
\begin{tabular}{c|ccccccc}
\hline
 Variables   &    $E$   &  $(123)$ &  $(321)$ & $(23)^*$ & $(13)^*$ &  $(12)^*$  \\
\hline
$r_{1}$     &   $r_{1}$   &  $r_{3}$   &  $r_{2}$ &   $r_{1}$   &  $r_{3}$   &  $r_{2}$  \\
$r_{2}$     &   $r_{2}$   &  $r_{1}$   &  $r_{3}$ &   $r_{3}$   &  $r_{2}$   &  $r_{1}$  \\
$r_{3}$     &   $r_{3}$   &  $r_{2}$   &  $r_{1}$ &   $r_{2}$   &  $r_{1}$   &  $r_{3}$  \\
$\alpha_{23}$   & $\alpha_{23}$   &$\alpha_{12}$ &$\alpha_{13}$   & $\alpha_{23}$   &$\alpha_{12}$ &$\alpha_{13}$       \\
$\alpha_{13}$   & $\alpha_{13}$   &$\alpha_{23}$ &$\alpha_{12}$   & $\alpha_{12}$   &$\alpha_{13}$ &$\alpha_{23}$       \\
$\alpha_{12}$   & $\alpha_{12}$   &$\alpha_{13}$ &$\alpha_{23}$   & $\alpha_{13}$   &$\alpha_{23}$ &$\alpha_{12}$       \\
\hline
\multicolumn{4}{c}{Characters} \\
\hline
Irrep $\Gamma$   &    $E$   & $(123)$ & $(23)^*$    \\
\hline
$A_1$       &   1   &  1    &   1    \\
$A_2$       &   1   &  1    &  -1    \\
$E  $       &   2   & -1    &   0   \\
\hline
\end{tabular}
\end{center}
\par
\label{t:C3v:r1r2r3}
\end{table}

%\Ans{As response to 1st referee suggestion to use a different program.}
The illustration below is based on the TROVE program again, however it should be transferable, at least in principle,  to
any method (i.e. any basis set, kinetic energy operator or potential energy function), as long as a similar choice of  vibrational coordinates and a product basis of 1D wavefunctions are used.
We choose the PH\3\ molecule and construct the symmetrized basis set in TROVE using a polyad number cutoff given by
$$
P  = 2 (n_1+n_2+n_3) + n_4 + n_5 + n_6 \le  P_{\rm max} = 10
$$
in conjunction with PES of \citet{15SoAlTe.PH3}. The 1D basis set functions are generated using the Numerov-Cooley procedure as described in Ref.~\citenum{15SoAlTe.PH3}, where the details on the kinetic energy expansion and the vibrational coordinates can also be found. In practice, we usually choose the maximal polyad number $P_{\rm max}$ in the order of 14--20   (see, for example, Refs.~\citenum{06YuCaTh.PH3,08OvThYu2.PH3,13SoYuTe.PH3,15SoAlTe.PH3}).

The basis set for subspace~1 (stretching) in this case contains only functions with  $n_1+n_2+n_3\le 5$ and $n_4=n_5=n_6=0$, while subspace 2 (bending) basis functions are constructed from the contributions
$n_4+n_5+n_6 \le 10$ and $n_1=n_2=n_3=0$. The first four variational eigenfunctions of $\hat{H}_{\rm str}^{(1)}$ read (where the shorthand notation $\ket{n_1,n_2,n_3}\equiv \ket{n_1}\ket{n_2}\ket{n_3}$ is used)
\begin{equation}
\label{e:Phi-XY3-before}
\begin{split}
\Psi_{1}^{(1)} &= 0.9997 \ket{0,0,0} -0.128\left( \ket{1,0,0} +  \ket{0,1,0} + \ket{0,0,1} \right)  + \ldots  \\
%\label{e:Phi-XY3-before-2}
\Psi_{2}^{(1)} &= −0.0223  \ket{0,0,0} -0.57689 \left( \ket{1,0,0} +  \ket{0,1,0} + \ket{0,0,1} \right) + \ldots   \\
\Psi_{3,1}^{(1)} &= 0.50667 \ket{0,0,1} -0.80753  \ket{0,1,0} +  0.30086 \ket{1,0,0}  + \ldots   \\
\Psi_{3,2}^{(1)} &= 0.63993 \ket{0,0,1}+ 0.11883  \ket{0,1,0} -  0.75875 \ket{1,0,0}  + \ldots
\end{split}
\end{equation}
corresponding to the energy term values $0.0$, 2317.86, 2328.28 and 2328.28~\cm, respectively, relative to the zero-point-energy (ZPE) of 5222.59~\cm\ (the actual coefficients are calculated in double precision).
In fact we find that high numerical accuracy is important for numerically reconstructing the symmetries of the eigensolutions.
The first two wavefunctions  $\Psi_{1}^{(1)}$ and $\Psi_{2}^{(1)}$ in Eq.~(\ref{e:Phi-XY3-before}) exhibit fully symmetric forms (to $\sim 10^{-15}$) and thus belong to $A_1$. The symmetry of the solutions $\Psi_{3,1}^{(1)}$ and $\Psi_{3,2}^{(1)}$ cannot be immediately recognized from their expansion coefficients, but the coinciding energy levels (within the numerical error of $\sim 10^{-15}$) indicate a degenerate solution, which for the case of \Cv{3}(M) can only mean the $E$ symmetry. For these two states we use the subscript notation $\lambda,n$ to refer to two degenerate components ($n=1,2$) of the degenerate state $\lambda=3$.

Thus, all four wavefunctions (as well as other solutions not shown here) appear readily symmetrized. However, guessing the degenerate $E$ symmetries based on the degeneracy of energies, which worked here, is not always reliable in actual numerical calculations, especially for high excitations and arbitrary potential functions with accidentally close energies (accidental resonances). In fact, some reduced Hamiltonian operators can lead to degenerate solutions of arbitrarily order, such as, for example, Harmonic oscillator or Rigid rotor Hamiltonians.
The $A_2$ states as single energy solutions can be also easily mixed up with $A_1$.

Finally, the degenerate components are usually come out of diagonalizations  as arbitrary degenerate mixtures. For example, the  eigenfunctions $\Psi_{3,1}^{(1)}$ and $\Psi_{3,2}^{(1)}$ from Eq.~(\ref{e:Phi-XY3-before}) were obtained using the eigen-solver DSYEV (LAPACK). As a result, they  do not necessarily obey the standard $E$-symmetry irreducible transformation rules: for example the ${\bf D}[(123)]$ transformation (which represents the $C_3$ rotation about the axis of symmetry) does not transform the two $E$-symmetry components  $(\Psi_{3,1}^{(1)},\Psi_{3,2}^{(1)})$ according to the transformation
$$
D[(123)] =
\left(\begin{array}{cc}
 \cos\epsilon  &    \sin\epsilon  \\
  -\sin\epsilon  &     \cos\epsilon
\end{array}
\right)
$$
with $\epsilon = 2\pi/3$, as one would expect.
In principle, such functions with randomly mixed components would still lead to a block-diagonal representation of the Hamiltonian matrix as in Eq.~\eqref{e:H-block}, and thus does not seem to be a problem. However having the functions $\Psi_{\lambda_i,n}^{(i)}$ which transform according with standard transformation rules is very useful for reducing the direct products of the corresponding irreducible representations. Therefore these randomly mixed degenerate components have to be further recast by a proper linear transformation to the standard form, which will be referred to as standard representations.

To conclude this section, the matrix symmetrization method based on reduced Hamiltonian operators can be efficiently used to produce a symmetry adapted basis set in fully numerical fashion. However,  the method does not tell which irreps these functions belong to and, consequently, which symmetry properties they have; besides, the degenerate components are mixed by an arbitrary orthogonal transformation which makes it difficult to use in subsequent calculations. This is where the second step of our symmetrization procedure, namely the symmetry sampling, comes in.

%\Que{18.- Line 52. P14. Please precise the meaning of ”standard E-symmetry”. By
%definition two orthonormal degenerate functions belonging to E carry an unitary rep-
%resentation D(R). ”They must be further reduced”. What does the word ”reduced”
%mean?. Change of basis? If this is the case the word is ill-used.} \Ans{Rephrased}
%\blue{\textit Since any linear combination of the degenerate eigenfunctions also a solution of the corresponding \schr\ equation %(\ref{e:XY3-red}), it is common to obtain arbitrary mixtures of them in numerical applications (see, e.g., %Ref.~\cite{11AlLeCa.CH4})}.

%\begin{figure}
%\begin{center}
%\caption{\label{f:XY3} The labeling of the nuclei and selected coordinates used for XY\3.}
% \leavevmode
%epsfxsize=10.0cm \epsfbox{XY3.eps}
%\end{center}
%\end{figure}

\section{Symmetry sampling of the eigenfunctions}
\label{s:sampling}

%\Que{19.- Section 4, as I understand is intended to establish an approach to generate
%the representation matrix D(R) given the degenerate functions spanning the irrep. However it is %not clear why this should be done. An anticipation of the need to calculate
%this is expected to be given at the beginning of the section.}
%\Ans{Try to better explain why the sampling is needed! Done?}

In this section we show how to reconstruct the symmetries $\Gamma_s$ of the eigenfunctions $\Psi_{\lambda_i,n}^{(i)}$ from  Eq.~\eqref{e:Hphi-reduced} by analyzing their transformation properties and also how to bring their degenerate components $n$ into the `standard' form.
Towards this end, we select a grid of $N_{\rm grid}^{(i)}$ instantaneous sampling geometries and use them to probe the values of the eigenfunctions $\Psi_{\lambda_i,n}^{(i)}$ as well as of their symmetry related images. That is,  for a given subspace $i$ and selected instantaneous geometries ${\bf Q}_k^{(i)}$ ($k=1\ldots N_{\rm grid}^{(i)}$), all symmetry related images  $R\, {\bf Q}_k^{(i)}$ are generated. These are used to reconstruct the values of the wavefunctions $\Psi_{\lambda_i,n}^{(i)}(R\,{\bf Q}_k^{(i)})$ at the new geometries, and to establish the transformation matrices ${\bf D}[R]$ for each operation $R$ of the group {\Group}. This is different from the more common practice of directly exploring the permutational properties of the wavefunctions. At this point it might appear that permuting the wavefunctions would be easier, at least for the case of $\ket{n_1}\ket{n_2}\ket{n_3}$ in our example of the rigid XY\3\ molecule. However, as will be shown below, applying the group transformations to the coordinates instead of the basis functions  provides a more general numerical approach applicable to complex cases when the permutation symmetry properties of the wavefunctions are not obvious.

%such as, e.g. $\xi_{a} = 1/\sqrt{6} (2 \delta \alpha_{23}-\delta \alpha_{12}-\delta \alpha_{13})$ and $\xi_{b} = 1/\sqrt{2} (\delta \alpha_{12}-\delta \alpha_{13})$ commonly used to represent the deformational modes of ammonia

Let us consider an $l_{\lambda}$-fold  degenerate eigenstate $\lambda$ with $l_{\lambda}$
eigenfunctions $\Psi^{(i)}_{\lambda,n}$ ($n=1,\ldots ,l_{\lambda}$) from a subspace $i$,
and define a grid of randomly  selected geometries ${\bf Q}_k^{(i)}$ ($k=1\ldots N_{\rm grid}^{(i)}$). We assume that the transformation properties of the coordinates from a given subspace with respect to $R$  are known at any specific point $k$. This can be expressed as:
\begin{equation}\label{e:Q:transform}
R \, {\bf Q}_k^{(i)} =  {\bf Q}{\p}_k^{(i)}
\end{equation}
with each subspace being independent from the others by definition.
%where $R$ \red{$R$ or $R$?} is a group operation including the unity operator $E$.
Under the assumption that the eigenfunctions $\Psi^{(i)}_{\lambda,n}({\bf Q}^{(i)})$ can be evaluated at any grid point $k$, i.e. at any instantaneous geometry ${\bf Q}_k^{(i)}$ including their symmetry related images ${\bf Q}{\p}_k^{(i)}$ (which is true for the TROVE program), we can evaluate
\begin{eqnarray}\label{e:phi:grid}
  \Psi^{(i)}_{\lambda,n}(k) &\equiv& \Psi^{(i)}_{\lambda,n}({\bf Q}_k^{(i)}),\\ \nonumber
  \Psi{\p}^{(i)}_{\lambda,n}(k) &\equiv& R\Psi^{(i)}_{\lambda,n}({\bf Q}_k^{(i)}) =  \Psi^{(i)}_{\lambda,n}({\bf Q}{\p}_k^{(i)}),
\end{eqnarray}
%\red{check if $R$ should be inverted here????}
where ${\bf Q}\p$ and $\Psi\p$ are the transformed coordinates and functions, respectively.
The eigenfunctions $\Psi^{(i)}_{\lambda,n}({\bf Q}_k^{(i)})$ and their symmetric images  $R\, \Psi^{(i)}_{\lambda,n}({\bf Q}_k^{(i)})$ are also related via the transformation matrices as given by:
\begin{equation}\label{e:represent-2}
  R\Psi^{(i)}_{\lambda,m}({\bf Q}_k^{(i)})  = \sum_{n=1}^{l_\lambda} D[R]_{mn} \Psi^{(i)}_{\lambda,n}({\bf Q}_k^{(i)}).
\end{equation}
It should be noted that we are using the convention by \citet{98BuJexx.method} to define the operations $R$ on the nuclear coordinates and functions. This convension is also referred to as passive (see, e.g., a detailed discussion by \citet{11AlLeCa.CH4}).
For instance, for the $E$-symmetry wavefunctions from Eq.~(\ref{e:Phi-XY3-before}), this expression reads
\begin{equation}\label{e:represent-XY3}
  \left(
  \begin{array}{c}
  \Psi{\p}_{3,1}^{(1)}(k) \\
  \Psi{\p}_{3,2}^{(1)}(k)
  \end{array}
  \right) =
  R
  \left(
  \begin{array}{c}
  \Psi_{3,1}^{(1)}(k) \\
  \Psi_{3,2}^{(1)}(k)
  \end{array}
  \right)
   =
\left(
  \begin{array}{cc}
  D_{11}[R] &  D_{12}[R] \\
  D_{21}[R] &  D_{22}[R]
  \end{array}
  \right)\cdot
  \left(
  \begin{array}{c}
  \Psi_{3,1}^{(1)}(k) \\
  \Psi_{3,2}^{(1)}(k)
  \end{array}
  \right).
\end{equation}
It should be noted that the linear system in Eq.~(\ref{e:represent-2}) does not impose the condition of unitariness of the solution. As a result the matrices $D[R]_{mn}$ can be non-orthogonal and must be orthogonalizied, for which the Gramm-Schmidt approach is employed.

Now by combining   Eq.~(\ref{e:phi:grid}) and Eq.~(\ref{e:represent-2}) we obtain
\begin{equation}\label{e:represent-combine}
  \sum_{n=1}^{l_\lambda} D[R]_{mn} \Psi_{\lambda,n}^{(i)}(k) = \Psi{\p}_{\lambda,m}^{(i)}(k).
\end{equation}
%where $n,m$ are components of the degenerate wavefunction $\Psi_{\lambda}^{\Gamma_s^{(i)}}$.

%\Que{20.- Line 52. P 16. ”li” is another notation out of the usual language of group
%theory. This is the dimension of the representation � as is explain and consequently I
%suggests to use n, in accordance to standard group theory language.} \Ans{We have stick to $l$ %instead of $n_{\Gamma}$ as in the book by Bunker and Jensen just because $n$ is already heavily %used.}
The $(l_\lambda)^2$ elements $D[R]_{mn}$ can be determined by solving Eq.~\eqref{e:represent-combine} as a system of $N_{\rm grid}^{(i)}$ linear equations ($k=1\ldots N_{\rm grid}^{(i)}$) of the type
\begin{equation}\label{e:lin}
\sum_n A_{kn}x_{n}^{(m)} =  b_k^{(m)}.
\end{equation}
Here the matrix elements $A_{kn}= \Psi_{\lambda,n}^{(i)}(k)$ and the vector-coefficients $b_k^{(m)} = \Psi{\p}_{\lambda,m}^{(i)}(k)$ are known, while  $x_n^{(m)} = D[R]_{nm}$ are the unknown quantities. Once the system of $N_{\rm grid}^{(i)}$ linear Equations \eqref{e:lin} is solved for each $R$ and all the $g$ transformation matrices ${\bf D}[R]$ are found ($g$ is the group order), we apply the standard projection operator approach \cite{98BuJexx.method} to generate the irreducible representations (see Section~\ref{s:projection}).

The number of degenerate reducible states $l_\lambda$ is simply taken as the number of states with the same energies. For non-degenerate wavefunctions ($l_\lambda=1$), the sampling procedure will always produce $D[R]_{1,1}=\chi[R]=\pm 1$.
Accidental degeneracies (e.g. $A_1$/$A_2$ with identical energies) are processed as if they were  normal degenerate components. In this case the resulting matrices are diagonal, $D[R]_{ii}=\pm 1$ and $D[R]_{i, j}=0$ ($i\ne j$).

At least $ N_{\rm grid}^{(i)} = (l_\lambda)^2$ grid points are required to define the linear system (or even less due to the unitary property of the transformation matrices). In practice, it is difficult to find a proper set of geometries  with all  values of $\Psi{\p}_{\lambda}^{(i)}(k)$ and $\Psi_{\lambda}^{(i)}(k)$ large enough to make the solution of the linear system numerically stable (i.e. with non-vanishing determinant). We therefore tend to select more  points ($N_{\rm grid}^{(i)} \gg (l_\lambda)^2$) and thus solve an over-determined linear system using the singular value decomposition method implemented in the LAPACK/DGELSS numerical procedure. We usually take $N_{\rm grid}^{(i)}$ = 10--200 geometries ${\bf Q}_k$ randomly distributed within the defined coordinate ranges around the equilibrium geometry of the molecule.
%We also make sure that none of the transformed geometries  ${\bf Q}{\p}_{k}$ (see %Eq.~\eqref{e:Q:transform}) leave the defined ranges.

This symmetrization procedure can be applied to any primitive functions provided their values can be calculated at any instantaneous geometry. For
example, the commonly used basis functions in TROVE are 1D eigensolutions of a
reduced 1D Hamiltonian determined using the Numerov-Cooley procedure and defined on an equidistant grid of geometries, typically of about 1000 points. In this case the values of the primitive functions $\phi_{n_k}(q_k)$ in Eq.~(\ref{e:basis-set}) are obtained by interpolation using the \textsc{polint} procedure \cite{07PrTeVe.method}. Other popular basis sets in TROVE are Harmonic
oscillator and  Rigid rotor wavefunctions, for which the details of the symmetrization procedure are presented below.

\section{Projection technique and symmetry classification}
\label{s:projection}

Due to the accidental  degeneracies and even more so due to the intrinsic degeneracies imposed by some reduced Hamiltonians (e.g. Hamiltonian of isotropic Harmonic oscillators), it is common to deal with degenerate solutions of Eq.~\eqref{e:Hphi-reduced} of high order, which can be much higher than that of the corresponding irreducible representations. For example, the Hamiltonian of the 2D Harmonic isotropic oscillator \citep{98BuJexx.method}
$$
 \hat{H}^{\rm 2D} = \frac{1}{2} \left[ P_a^2 + P_b^2 + \lambda (Q_a^2 + Q_b^2) \right]
$$
has the eigenvalues
$$
E_{v_a,v_b}^{\rm 2D} = \hbar \sqrt{\lambda} \left[ v_a + v_b +1 \right],
$$
which are $(v_a + v_b+1)$-fold degenerate. As it was discussed above, our numerical symmetrization approach often leads to arbitrarily mixed degenerate representations, which need to be further transformed to the standard orthogonal form. In the following we show how to use the standard projection technique to symmetrize such general cases in a fully numerical fashion.

%\Que{21.- Section 5 starts with a reducible representation. I suppose this situation is
%considered when accidental degeneracy appear. But if this is the situation it is possible
%that multiplicity of representation appear. This situation is not discussed. Please
%comment about it.}
%\Ans{Add discussion of multiple representations }

In order to reduce a representation $\Gamma_{\rm red}$ to its irreducible components
\begin{equation}\label{e:Gamma:a}
  \Gamma_{\rm red} = a_1 \Gamma_1 \oplus a_2 \Gamma_2 \oplus  a_3 \Gamma_3 \oplus  \cdots \oplus  a_h \Gamma_h,
\end{equation}
the first step is to use the characters $\chi[R]$ of the reducible representation as traces of the transformation matrices $D[R]_{mn}$:
$$
\chi[R] = \sum_{n} D[R]_{nn}
$$
and find the number of irreducible representations  $a_{s}$ (reduction coefficients)  for each irreps $\Gamma \in $ \Group\ as given by:
\begin{equation}\label{e:reduction}
  a_s = \frac{1}{g} \sum_{R} \chi[R]^* \, \chi^{\Gamma_s}[R].
\end{equation}
Remember that $g$ is the order of the group, $R$ runs over all the elements of the group and $\chi^{\Gamma}[R]$ are the group characters. To ensure the numerical stability of the symmetrization we usually check if these reduction coefficients are (i) integral and (ii) satisfy the reduction relations \cite{98BuJexx.method}
\begin{equation}\label{e:charcters:relation}
  \chi[R] = \sum_{s} a_s \chi^{\Gamma_s}[R], \quad {\rm and} \quad \sum_{R} \left| \chi^{\Gamma_s}[R] \right|^2 = g.
\end{equation}
If these conditions are not fulfilled (within some numerical thresholds, typically $10^{-3}$), the grid points are re-selected and the transformation matrices are re-built.

%After this the transformation matrices are used in the reduction described above.

In principle a projection onto a non-degenerate irrep $\Gamma_s$
can be generated  by the operator \cite{98BuJexx.method}:
\begin{equation}\label{e:projection}
  P^{\Gamma_s}  = \frac{1}{g} \sum_{R} \chi^{\Gamma_s} [R]^{*} R.
\end{equation}
However, a non-degenerate function $\Psi_{\lambda}^{(i)}$ obtained using Eq.~(\ref{e:Hphi-reduced}) should already transform irreducibly as $\Gamma_s$, and therefore this projector $P^{\Gamma_s}$ ($\Gamma_s \in$ \Group) is not needed. For example, applying the reduction from  Eq.~(\ref{e:reduction}) to the first two non-degenerate wavefunctions ($\Psi_{1}^{(1)}$ and $\Psi_{2}^{(1)}$) in Eq.~(\ref{e:Phi-XY3-before}) will give $a_{A_1}=1$, $a_{A_2}=a_{E} = 0 $, which unambiguously defines their symmetries.

Degenerate solutions require special care. For the sake of generality let us assume that degeneracy of the reducible solution $l_\lambda$ can be higher than that of the irreducible representations $l_s$. The degenerate wavefunctions (both accidentally and intrinsically) can be selected simply based on the coincidence of energies within a specified threshold (usually 0.001~\cm). The corresponding transformation $l_\lambda\times l_\lambda$ matrices ${\bf D}[R]$ are constructed using the sampling procedure of Eq.~(\ref{e:represent-2}) and then symmetrized with Eq.~(\ref{e:reduction}) giving  the reduction coefficients of irreps $\Gamma_s$.

In cases of multiple degenerate states ($l_\lambda>1$), the following transfer operator is used\citet{98BuJexx.method}
\begin{equation}\label{e:transfer}
  P_{mn}^{\Gamma_s} = \frac{l_s}{g} \sum_{R} D^{\Gamma_s}[R]^{*}_{mn} R,
\end{equation}
where ${\bf D}^{\Gamma_s}[R]$ is an  irreducible orthogonal  transformation matrix of $\Gamma_s$ for an  operation $R$, and $l_s$ is the dimension (degeneracy) of $\Gamma_s$. Following the symmetrization protocol \cite{98BuJexx.method} and using the reducible  $D^{\Gamma}[R]_{mn}$ from Eq.~\eqref{e:represent-2} as a representation of $R$, the $m$th component of the irreducible wavefunction $\Psi_{\lambda,m}^{\Gamma_s}$ is obtained by acting $P_{mm}^{\Gamma_s}$ (diagonal element) on $\Psi_{\lambda,n}^{\Gamma_{\rm red}}$. Here we distinguish the reducible and irreducible representations  by using the superscripts  $\Gamma_{\rm red}$  and and $\Gamma_s$, respectively.
The off-diagonal elements of the transfer operator $P_{m,n}^{\Gamma_s}$  are then used to recover other components of $\Psi_{\lambda,n}^{\Gamma_s}$.

%\red{rephrase}
It should be noted that degenerate solutions $\Psi_{\lambda,n}^{\Gamma_{\rm red}}$  in general can span more than one representation. Besides the projected vectors are not automatically orthogonal. Therefore the symmetry classification procedure must include the following steps: (i)
The projector $P_{mm}^{\Gamma_s}$ ($m=1\ldots l_\lambda$) is applied to $\Psi_{\lambda,p}^{\Gamma_{\rm red}}$ to form a trial irreducible solution $\Psi_{\lambda,m}^{\Gamma_s}$, which is then (ii) checked against already found functions $\Psi_{\lambda,n}^{\Gamma_s}$ ($n<m$). The trial function is then either rejected (if it is already present in the set) or (iii) orthogonalized to this set using the Gramm-Scmidt orthogonalization technique. This procedure is repeated until all $a_s$ irreducible solutions are found.

% and the operator given by Eq.~\eqref{e:projection} can in principle
%project to the right irrep $\Gamma$ (i.e. with the right characters), but does not guarantee that the degenerate %$\Gamma_s^{(i)}$ will have standard transformation properties.

%\red{CHECK WHY SIDE THERE IS NO PROJECTOR HERE!}
%\begin{equation}\label{e:represe:non-symm}
%  R \phi_{\lambda,m}^{\Gamma_s^{(i)}} =  \sum_{n} D^{\Gamma}[R]_{mn} \phi_{\lambda,n}^{\Gamma_s^{(i)}}.
%\end{equation}
%and applying the diagonal element of the transfer operator $P_{mm}^{\Gamma_s^{(i)}}$
%(called projection operator \cite{98BuJexx.method})
%to $\Psi_{\lambda,n}^{\Gamma}$ we obtain the $m$th-component function $\Psi_{\lambda,m}^{\Gamma_s^{(i)}}$ (where we %distinguish them by the superscripts  $\Gamma$ \red{\bf(reducible?)} (mixed) and $\Gamma_s^{(i)}$ \red{\bf(irreducible?)} %(standard)), while the non-diagonal elements of $P_{mp}^{\Gamma_s^{(i)}}$ can recover its $p$th component %$\Psi_{\lambda,p}^{\Gamma_s^{(i)}}$.

%The projection technique is applied to the set of degenerate functions $\Psi_{\lambda,n}^{\Gamma}$ ($n=1\ldots s$) as follows.

\subsection{Tetratomics of the XY$_3$-type, \Cvvv-symmetry (Continued)}
\label{s:XY3:cont}

Let us now return to the example above.
Choosing 40 points and applying our sampling approach to the degenerate state $\Psi_{3,n}^{(1)}$ in Eq.~(\ref{e:Phi-XY3-before}) for all six  \Cv{3}(M)
%permutation operations to $r_1,r_2,r_3$ as
group operations listed
in Table~\ref{t:C3v:r1r2r3},  the following transformation matrices
%in the reducible representation $\Gamma$
were determined:
%\Que{22.- In Section 5.1, why it is necessary to list the six representation matrices? It is
%enough with the generators, is not it?} \Ans{Give only one per class? Or explain why we show all %six, just for illustrative purposes?}
\begin{eqnarray}
 {\bf D}[E] =
\left(\begin{array}{cc}
   1.0000  &      0.0000   \\
   0.0000  &      1.0000
\end{array}
\right),  & \;
 {\bf D}[(123)] =
\left(\begin{array}{cc}
 -0.5000  &      0.8660   \\
 -0.8660  &     -0.5000
\end{array}
\right), \\
 {\bf D}[(321)] =
\left(\begin{array}{cc}
 -0.5000  &     -0.8660   \\
  0.8660  &     -0.5000
\end{array}
\right), & \;
 {\bf D}[(23)^*] =
\left(\begin{array}{cc}
 -0.5813  &     -0.8137   \\
 -0.8137  &      0.5813
\end{array}
\right), \\
 {\bf D}[(13)^*] =
\left(\begin{array}{cc}
  0.9953  &     -0.0965   \\
 -0.0965  &     -0.9953
\end{array}
\right),
& \;
 {\bf D}[(12)^*] =
\left(\begin{array}{cc}
 -0.4141  &      0.9102   \\
  0.9102  &      0.4141
\end{array}
\right).
\end{eqnarray}
%0.999999999999857706415677209103402
In principle only three matrices are unique, but TROVE currently computes matrices for all representations and does not take the advantage of generators.
The characters $\chi^{\Gamma}[R] $ of these transformations are $2.0$, $-1.0$, and $0.0$ ($\pm 10^{-12}$), which in conjunction with Eq.~(\ref{e:reduction}) leads to the following reduction coefficients $a^{E} = 1 $ and $a^{A_1}=a^{A_2}=0$ ($\pm 10^{-12}$)  as expected for a doubly-degenerate solution.

Using the transformation matrices ${\bf D}[R]$ together with Eq.~\eqref{e:transfer}, we build a projection operator $P_{11}^E$ and apply it to the degenerate components $\Psi_{a} = \Psi_{3,1}^{(1)}$ and $\Psi_{b} = \Psi_{3,2}^{(1)}$ to obtain a trial vector:
$$
\tilde\Psi_a =   0.135862903\, \Psi_a  -0.342642926\, \Psi_b,
$$
which after normalization becomes
$$
\tilde\Psi_a =   0.368595853\, \Psi_a -0.929589747\, \Psi_b.
$$
The second component $\tilde\Psi_b$ is found by applying the transfer operator in Eq.~\eqref{e:transfer}:
%this we use the transfer-operator \red{PROJECTOR?} matrices defined as follows \cite{98BuJexx.method}
%\begin{equation}\label{e:transfer}
%  P_{mt}^{\Gamma_s^{(i)}} = \frac{n_{\Gamma_s^{(i)}}}{h} \sum_{R} D^{\Gamma_s^{(i)}}[R]^{*}_{mt} R,
%\end{equation}
$$
\tilde\Psi_b = \frac{2}{6} \sum_{R} D^{\Gamma_s}[R]_{12}^* \tilde\Psi_a
$$
which when normalized reads
$\tilde\Psi_b   = 0.92958974\, \Psi_a + 0.368595853\, \Psi_b$.

Finally, by applying the transformation vectors to the original (reducible) representation $\{\Psi_a,\Psi_b\}$ from Eq.~(\ref{e:Phi-XY3-before}) we obtain
\begin{eqnarray}
\label{e:Phi-XY3-after-1}
\tilde\Psi_{3,1}^{(1)} = \tilde\Psi_{E_a}^{(1)} &=&  -\frac{1}{\sqrt{6}} \left( \, \ket{0}\ket{0}\ket{1} +  \ket{0}\ket{1}\ket{0} - 2 \, \ket{1}\ket{0}\ket{0} \, \right) + \ldots \\
\label{e:Phi-XY3-after-2}
\tilde\Psi_{3,2}^{(1)} = \tilde\Psi_{E_b}^{(1)} &=& \phantom{-} \frac{1}{\sqrt{2}}   \left(\, \ket{0}\ket{0}\ket{1} - \ket{0}\ket{1}\ket{0} \, \right) + \ldots
\end{eqnarray}
which is the well-known form that transforms according to the standard $E$-symmetry representations of \Cv{3} (see Ref.~\citenum{03CeSpxx} for example). The expansion coefficients in Eqs.~(\ref{e:Phi-XY3-after-1},\ref{e:Phi-XY3-after-2}) are defined within a numerical error of $10^{-14}$.
%States with $\lambda_1 = 3$ and  $\lambda_2 = 3$ in  Table~\ref{t:energy:XY3}  are degenerate and described by two eigenfunctions given by Eqs.~(\ref{e:Phi-XY3-after-1},\ref{e:Phi-XY3-after-2}). The eigenstates can be also assigned with quantum numbers based on the largest coefficient contribution from the basis set expansion, also shown in Table~\ref{t:energy:XY3}. Here we use $v_1, v_2, v_3, v_4, v_5$, and $v_6$ as the corresponding excitation quanta. At this point it is relatively trivial to match TROVE `local' mode quantum numbers with the `normal' mode quantum numbers, also shown in the spectroscopic notation. See also Ref.~\cite{08OvThYu.PH3} (Table~II) for the extended set of TROVE vibrational energies and assignments of PH$_3$.

We can check if the new vectors transform correctly, as in this case, i.e. according to the standard irreducible matrices ${\bf D}^{\Gamma_s}[R]$ as follows
$$
R \Psi_{\lambda,m}^{(i),\Gamma_s} = \sum_{n} D^{\Gamma_s}[R]_{m,n} \Psi_{\lambda,n}^{(i),\Gamma_s} .
$$
As mentioned above, if the projection operator $P_{mm}^{\Gamma_s}$ does not lead to a correct or independent combination, we would try a different component of $P_{mm}^{\Gamma_s}$ until the correct solution is found (which is guaranteed).

With this procedure, symmetries of all eigenstates can be easily reconstructed. For the basis set  $P\le P_{\rm max} = 10$ in this example (see Section~\ref{s:XY3}), we computed 38 stretching $\Psi_{\lambda_1}^{(1)}$ and 192 bending  $\Psi_{\lambda_2}^{(2)}$ eigenfunctions, with the  symmetries and energies of the first 3 from each subspace shown in Table~\ref{t:energy:XY3}.

\begin{table}
\begin{center}
\caption{\label{t:energy:XY3}
Term energies $\tilde{E}_{\lambda}$ (in cm$^{-1}$), symmetries $\Gamma$, degeneracies $l_\lambda$, TROVE assignments $(v_i)$, and normal-mode assignments $(\nu_i)$ of the eigensolutions for each subspace $i=$1, 2.}
\begin{tabular}{ccrccccccccc}
\hline
\hline
Subspace $i$  &   $\lambda_i$ &  $\tilde{E}_{\lambda_i}$ &  $l_{\lambda_i}$  &$  \Gamma$ &$  v_1 $&$  v_2  $&$ v_3  $&$ v_4  $&$ v_5 $&$ v_6$ & State   \\
\hline
     1  &     1  &          0.0  &    1    &$    A_1    $&$    0 $&$     0 $&$    0 $&$    0 $&$   0 $&$    0 $& 0  \\
     1  &     2  &   2320.0950  &    1    &$    A_1    $&$    0 $&$     1 $&$    0 $&$    0 $&$   0 $&$    0 $& $\nu_1$ \\
     1  &     3  &  2329.6856  &    2    &$     E     $&$    1 $&$     0 $&$    0 $&$    0 $&$   0 $&$    0 $& $\nu_3$ \\
     2  &     1  &          0.0  &    1    &$    A_1    $&$    0 $&$     0 $&$    0 $&$    0 $&$   0 $&$    0 $& 0 \\
     2  &     2  &  1013.7488  &    1    &$    A_1    $&$    0 $&$     0 $&$    0 $&$    1 $&$   0 $&$    0 $& $\nu_2$ \\
     2  &     3  &  1121.4813  &    2    &$     E     $&$    0 $&$     0 $&$    0 $&$    1 $&$   0 $&$    0 $& $\nu_4$ \\
\hline
\hline
\end{tabular}
\end{center}
\end{table}

%In this table we also show the value of the largest expansion coefficients $|C_{i}|^{(k),\lambda_k}$ ($k1,2$), which we routinely use in order to assign quantum numbers to the corresponding states, see, for example, Ref.~\cite{PH3-paper?-assignement?}.

Once all symmetry adapted eigenfunctions for each subspace $i=1,2$ are found, the final vibrational basis set is formed as a direct product
$$
\Psi_{\lambda_1,\lambda_2}^{\Gamma_1,\Gamma_2} = \Psi_{\lambda_1}^{(1),\Gamma_1} \otimes \Psi_{\lambda_2}^{(2),\Gamma_2}  ,
$$
which is not irreducible and has to be further symmetrized. We use the same projection/transfer operator approach described above (and even  the same numerical subroutine) by Eqs.~(\ref{e:projection}, \ref{e:transfer}).  The required transformation matrices are obtained as products of the standard irreducible  transformation matrices
$$
{\bf D}^{\Gamma_1,\Gamma_2}[R] = {\bf D}^{\Gamma_1}[R]\,  {\bf D}^{\Gamma_2}[R]
$$
which are well known and also programmed in TROVE for most symmetry groups. Using standard transformation matrices is numerically more stable compared to the procedure based on the matrices ${\bf D}[R]$  evaluated directly as solutions of Eq.~(\ref{e:represent-combine}). This is exhibited in significantly smaller errors in the computed coefficients $a_i$, which are very close to being integral.
%In Table~\ref{t:xxx} we show the irreducible representations of the products of the stretching (1) and bending (2) subspaces for our %example with $P_{\rm max}$ =1. According with the polyad contraction scheme, all combinations with $v_1+v_2+v_3+v_4+v_5+v_6>1$ are %excluded. \red{WE DON'T HAVE THIS YET. DO WE NEED IT?}

To illustrate this point, it is informative to look at the product of two degenerate functions $\Psi_{3}^{(1),E} \otimes \Psi_{3}^{(2),E} $ as an example (see Table~\ref{t:energy:XY3}).
%\Que{23.- Line 12. P 23. The notation is not formal, component coefficients are missing.}
%For example, the combination $\Psi_{3}^{E} \otimes \Psi_{3}^{E} $ is reduced as follows.
The four components of the product $\Psi_{3,n}^{(1),E} \Psi_{3,m}^{(2),E}$ ($n,m$=1,2) transform as a direct product of two $E$-representation matrices \cite{98BuJexx.method}
$$
{\bf D}^{\Gamma_s}[R] =  {\bf D}^{E}[R] \otimes {\bf D}^{E}[R].
$$
%Here $E_a$ and $E_b$ denote the two components of the 2D irreducible representation $E$.
The characters are defined by
$$
\chi^{\Gamma_s}[R] = (\chi^{E})^2
$$
and give 4, 1, and 0 for $E$, $(123)$, and $(12)^*$, respectively, which is the standard textbook
%\Que{24.- Line 30. P 23. Sign x must be Oplus}
example of a reduction of the $E\otimes E$ product (see, for instance, Ref.~\citenum{98BuJexx.method}). The reduction coefficients as obtained from Eq.~\eqref{e:reduction} are 1, 1, 1 for $A_1$, $A_2$, and $E$, i.e.
$$
E\otimes E= A_1 \oplus A_2 \oplus E.
$$
The irreducible representations determined using the numerical approach described above are
\begin{eqnarray}
\Psi_{3,3}^{A_1} &=& \frac{1}{\sqrt{2}} \left[  \Psi_{3,3}^{(E_a,E_a)} + \Psi_{3,3}^{(E_b,E_b)} \right] ,\\
\Psi_{3,3}^{A_2} &=& \frac{1}{\sqrt{2}} \left[  \Psi_{3,3}^{(E_a,E_b)} - \Psi_{3,3}^{(E_b,E_a)} \right], \\
\Psi_{3,3}^{E_a} &=& \frac{1}{\sqrt{2}} \left[  \Psi_{3,3}^{(E_a,E_a)} + \Psi_{3,3}^{(E_b,E_b)} \right], \\
\Psi_{3,3}^{E_b} &=& \frac{1}{\sqrt{2}} \left[  \Psi_{3,3}^{(E_a,E_b)} - \Psi_{3,3}^{(E_b,E_a)} \right],
\end{eqnarray}
where the corresponding expansion coefficients $\pm 1/\sqrt{2}$  are obtained numerically with double precision accuracy.

This completes the PH\3\ example as well as the description of the TROVE numerical symmetrization procedure. The approach is very robust and is applicable to any product-type basis sets constructed from 1D functions provided the transformation rules for the coordinates are known. The most time-consuming part of our numerical implementation is the sampling procedure which relies on the random selection of points and can occasionally lead to poor solutions of Eq.~(\ref{e:represent-combine}) for the transformation matrices.  Usually the calculations are quick (seconds) but  sometimes they can take hours (remember this is a basis set initialization part which has to be done only once).

%\blue{It is only for a 4D isotropic oscillator \schr\ equation as a reduced eigen-problem (Eq.~\eqref{e:Hphi-reduced}) %with the degeneracy of the XXX order (the cardinal number $N=YY$) the symmetrization becomes impractical.}

%The product of $\phi_{\lambda_1}^{(1)}$ and $\phi_{\lambda_2}^{(2)}$ forms the vibrational basis set, which is then also symmetrized using the projection operators as described above. The corresponding transformational matrices $D[R]_{mm\p}$ are assumed to be agree with the standard form exactly, where some level of averaging is involved. Therefore the irreducible representation coefficients $a_i$, which are treated as real values, all obtained very close to be integral.

\subsection{An XY\3\ molecule of \Dh{3}\ symmetry:  Degenerate  and redundant coordinates}
\label{s:XY3:degen}

Let us consider a more complicated example of coordinate choice, where some of the coordinates transform according to two-fold irreducible representations.
Such coordinates are commonly used to describe the vibrations of non-rigid molecules. For example,  the nuclear coordinates of ammonia can be defined as:
\begin{eqnarray}
q_{1} & = &  \Delta r_1 \\
q_{2} & = &  \Delta r_2 \\
q_{3} & = &  \Delta r_3 \\
\label{e:q4:NH3}
q_{4} & = &   \frac{1}{\sqrt{6}} \left[ 2 \Delta \alpha_{23} - \Delta \alpha_{12} - \Delta \alpha_{13}  \right] \\
\label{e:q5:NH3}
q_{5} & = &   \frac{1}{\sqrt{2}} \left[ \Delta \alpha_{12} - \Delta \alpha_{13}  \right] \\ q_{6} & = &  \tau.
\end{eqnarray}
Here, $r_1$, $r_2$, $r_3$ are the bond lengths, $\alpha_{23}$, $\alpha_{12}$, and $\alpha_{13}$ are the interbond angles and $\tau$ is the inversion `umbrella' coordinate measuring the angle between a bond and the trisector (see Ref.~\citenum{02LeHaSt.NH3} for example).

In this case the vibrational modes span three subspaces, stretching $\{ q_1, q_2, q_3\}$, bending $\{ q_4, q_5\}$, and inversion $\{ q_6\}$, which transform independently.
 %(see Table~\ref{t:D3h-r-S-delta} in Appendix).
The symmetry properties of the two bending modes
special compared to those of the stretching and inversion modes, where the effect of the symmetry operations on the latter is just a permutation
$$
R q_i = q_j, \;\; i,j = 1,2,3,
$$
or a change of sign,
$$
R q_6 =  \pm q_6.
$$
Whereas for the two asymmetric bending coordinates $q_4$ and $q_5$ (which are based on three redundant coordinates $\alpha_{23}$, $\alpha_{12}$, and $\alpha_{13}$)
are mixed by the degenerate $E$-symmetry transformations:
$$
R \left(
\begin{array}{c}
q_4 \\
q_5
\end{array}
\right) = \left(
\begin{array}{c}
D^{E}[R]_{11}\, q_4  +  D^{E}[R]_{12}\, q_5 \\
D^{E}[R]_{21}\, q_4  +  D^{E}[R]_{22}\, q_5
\end{array}
\right).
$$

The product-type primitive basis set for NH$_3$ ($J=0$) is %(see, for example, Ref.~\cite{09YuBaYa})
%\Que{25.- In Eq.(49) the notation is not clear given Eq.(2). Where are the quantum
%numbers n?}
\begin{equation}
\label{e:prim-basis-D3h}
  \phi_{\nu}({\bf Q}) =
\phi_{n_1}(q_1) \phi_{n_2}(q_2) \phi_{n_3}(q_3) \phi_{n_4}(q_4) \phi_{n_5}(q_5) \phi_{n_6}(q_6),
\end{equation}
where $\phi_{n_k}(q_k)\equiv|n_k\rangle$ ($k=1\ldots 6$) are 1D primitive basis functions.
Due to the 2D character of the transformations of $q_4$ and $q_5$, the primitive bending functions $\phi_{n_4}(q_4)$ and $\phi_{n_5}(q_5)$ do not follow  simple permutation symmetric properties. For example, by applying the $(123)$ permutation to the product $\phi_{n_4}(q_4) \phi_{n_5}(q_5)$ we get:
$$
 (123) \phi_{n_4}(q_4) \phi_{n_5}(q_5) =
 \phi_{n_4}( -\frac{1}{2} q_4 + \frac{\sqrt{3}}{2} q_5 )
 \phi_{n_5}(  -\frac{\sqrt{3}}{2} q_4 -  \frac{1}{2} q_5 ),
$$
 which cannot be expressed in terms of products of $\phi_{n_4}(q_4)$ and $\phi_{n_5}(q_5)$ only. Strictly speaking, an infinite primitive basis set expansion in terms of $\phi_{n_4}(q_4) \phi_{n_5}(q_5)$ is required to represent $R \phi_{n_4}(q_4) \phi_{n_5}(q_5)$ exactly, except for the special case of Harmonic oscillator functions (see Section~\ref{s:Harmonic}).
 In practice, we use expansions large enough to converge the symmetrization error below the defined threshold of $10^{-14}$. Unlike the two above examples of rigid molecules, the lack of the permutation character of the product-type basis set $\phi_{n_1,\ldots, n_6}({\bf Q})$ in Eq.~\eqref{e:prim-basis-D3h} also prevents its symmetrization using the transformation properties of the functions. However, our approach is based on the transformation properties of the coordinates ${\bf Q}$, not functions, which allows a symmetry adapted representation to be constructed even in this case.

The first step is to build three reduced Hamiltonian operators for each $i=1,2,3$ subspace of coordinates
\begin{eqnarray}
\label{e:HXY3:D3h:1}
  \hat{H}^{(1)}(q_1,q_2,q_3) &=& \bra{0_4} \bra{0_5} \bra{0_6} \hat{H} \ket{0_6} \ket{0_5} \ket{0_4}, \\
\label{e:HXY3:D3h:2}
  \hat{H}^{(2)}(q_4,q_5) &=& \bra{0_1} \bra{0_2} \bra{0_3} \bra{0_6} \hat{H} \ket{0_6} \ket{0_3} \ket{0_2} \ket{0_1}, \\
\label{e:HXY3:D3h:3}
  \hat{H}^{(3)}(q_6) &=& \bra{0_1} \bra{0_2} \bra{0_3} \bra{0_4} \bra{0_5} \hat{H} \ket{0_5} \ket{0_4} \ket{0_3} \ket{0_2} \ket{0_1}
\end{eqnarray}
and solve the corresponding eigenvalue problems
\begin{equation}
\label{e:D34-eigen}
  \hat{H}^{(i)}({\bf Q}^{(i)}) \Psi_{\lambda_i}^{(i)}({\bf Q}^{(i)}) = E_{\lambda_i} \Psi_{\lambda_i}^{(i)}({\bf Q}^{(i)}).
\end{equation}
As discussed above, we expect all eigenvectors of Eq.~\eqref{e:D34-eigen} to transform according to the irreducible representations $\Gamma_s^{(i)}$ of \Dh{3}(M) despite the non-permutative character of the bending primitive functions. It should be noted that in practical calculations, employing a finite basis set affects the accuracy with which the irreducible character of the eigenfunctions can be determined,
which is particularly true for high vibrational excitations $n_k$.

To illustrate this, let us consider a generic variational calculation of several lower eigenstates
for ammonia. Here we use the PES from Ref.~\citenum{09YuBaYa} and the primitive basis set defined by a polyad number $P$ of
%%%%%%%%%%%%%%%%%%%%%%%%%%%%%%%%%%
\begin{equation}\label{e:polyad-2}
    P =  2 (n_1 + n_2 +n_3) + n_{4} + n_{5} + n_{6}/2 \le P_{\rm max} = 28.
\end{equation}
The primitive basis functions $\phi_{n_k}(q_k)$ ($k=1..6$) are obtained as eigensolutions of the corresponding 1D reduced Hamiltonians using the Numerov-Cooley technique \cite{23Nuxxxx.method,61Coxxxx.method}
%%%%%%%%%%%%%%%%%%%%%%%%%%%%%%%%%%
with a computational setup as described in  Ref.~\citenum{09YuBaYa}.
The solution of the reduced stretching problem in Eq.~\eqref{e:HXY3:D3h:1} is equivalent to the example of the rigid XY\3 example detailed above  (see Eqs.~(\ref{e:Phi-XY3-before})) and is not discussed further. The first three solutions of the bending reduced problem in Eq.~\eqref{e:HXY3:D3h:1} are given by (where $|n_4,n_5\rangle\equiv|n_4\rangle|n_5\rangle$)
\begin{eqnarray}
\label{e:Phi-D3h-before-1}
\Psi_{1}^{(2)}   &=& \phantom{-} 0.99995 \, \ket{0,0} -0.00741 \, \left( \ket{2,0} +  \ket{0,2}  \right)  + \ldots  \\
\label{e:Phi-D3h-before-2}
\Psi_{2,1}^{(2)} &=& \phantom{-}0.00377 \, \ket{0,1} + 0.99988 \, \ket{1,0}  + \ldots   \\
\Psi_{2,2}^{(2)} &=&-0.99988 \, \ket{0,1} + 0.00377 \,  \ket{1,0}  + \ldots  .
\end{eqnarray}
with the energy term values of 0.0, 1679.6324 and 1679.6324~\cm\ relative to the ZPE = 1953.7381~\cm.

The wavefunctions $\Psi_{2,1}^{(2)}$ and $\Psi_{2,2}^{(2)}$ are recognized as degenerate ($\lambda_2 = 2$) due to their very similar energies (we use a threshold of $10^{-6}$~\cm)
and  should be processed together at the symmetrization step. The transformation matrices ${\bf D}[R]$ are determined by sampling the eigenfunctions on a grid  of 40 points to give the  reduction coefficients $a_i$ = 0, 0, 1, 0, 0 and 0 ($\pm 10^{-12}$) for $A_1\p$, $A_2\p$, $E\p$,$A_1\pp$, $A_2\pp$ and $E\pp$, respectively. The projection operator procedure leads to the symmetrized combinations given by
\begin{eqnarray}
 \label{e:Phi-D3h-after-1}
 \Psi_{2,1}^{(2)} &=&-0.99989 \, \ket{1,0} -0.00741 \, \ket{1,2} -0.01283 \, \ket{3,0} +  \ldots   \\
 \label{e:Phi-D3h-after-2}
 \Psi_{2,2}^{(2)} &=&-0.99989 \, \ket{0,1} + 0.00741 \, \ket{2,1} +0.01283 \, \ket{0,3} +  \ldots
\end{eqnarray}

Reducing the basis set to $P_{\rm max} = 2$, i.e. taking only $n_4,n_5 \le 1$, leads to similar solutions but with  larger errors of about $10^{-8}$ for $a_i$, which is still rather small in this case.
%Apparently the corresponding basis functions are very close to the Harmonic oscillators which can be exactly re-expanded.  It %appears that the lower states solutions are well behaved and can be easily symmetrized.
However, the wavefunctions corresponding to higher excitations will introduce larger errors and will require more basis functions for accurate symmetrization. We use a threshold of $10^{-3}$--$10^{-4}$ for reduction coefficients $a_i$  to control the symmetrization procedure: the program will accept solutions if $a_i$ differ from an integer by less than this value.

As a final and conclusive test, TROVE also checks the matrix elements of the total Hamiltonian $\hat{H}$ between different symmetries, which should be vanishingly small to allow a block-diagonal form of the Hamiltonian matrix. TROVE uses an acceptance threshold of $10^{-3}$--$10^{-4}$~\cm\ to control the quality of the symmetrization procedure. Failure to pass this test (usually small errors) indicates that the basis set is not large enough for an accurate symmetrization. Critical failure  (huge errors) usually means problems with the model (e.g. in the potential energy function, coordinate transformation relations, kinetic energy operator, definition of the molecular equilibrium structure etc).

%\Que{26.-The example presented in Section 5.2 presents redundant coordinates in the
%subspace for the bends. Is there any consequence of this fact in this approach?}
This example is a good illustration of how the redundant coordinates can be incorporated into a product-type basis of 1D functions. The redundant vibrational coordinates are very common, for example they appear as part of multi-dimensional symmetrically adapted coordinates.
The typical example are the bending modes used to represent vibrational modes of ammonia, Eqs.~(\ref{e:q4:NH3},\ref{e:q5:NH3}) or methane (see, for example, Ref.~\citenum{97Haxxxx.CH4}).
The TROVE symmetrization can still handle this situation even at a cost of a larger basis set.
As it will be shown in the next section, the Harmonic oscillator basis functions have the property of their products to form symmetrized
combinations from a finite size basis of functions, which holds also for the case of the redundant coordinates.

\subsection{Harmonic oscillator basis sets}
\label{s:Harmonic}

Our most common choice of the primitive basis set is based on the Numerov-Cooley approach, where 1D functions are generated numerically on a large grid of 1000--5000 equidistantly placed points by solving a set of 1D reduced Hamiltonian problems for each mode. This provides a compact basis set optimized for a specific problem. However, as was discussed in the previous section, some types of degenerate coordinates require large expansions in terms of products of 1D functions for accurate symmetrization. A very simple work-around of this problem is to use 1D Harmonic oscillators as a basis set. The (degenerate) Harmonic oscillators have a unique property: one can always build an isotropic Harmonic oscillator with proper symmetric properties as a finite sum of products of 1D Harmonic oscillators  $\phi_{n_i}^{\rm HO}(q_i)$ (see, for example, Ref.~\citenum{98BuJexx.method}). This is also valid for higher milti-fold degeneracies. As an illustration, in the Appendix we show how to construct a 2D symmetrized basis set using 1D Harmonic oscillator functions to represent the asymmetric bending modes of the ammonia molecule using our symmetrization procedure.
In fact this illustration can be reproduced  without the TROVE program as it is solely based on the properties of the Harmonic wavefunctions.
This makes up a good toy example to try our symmetrization approach without having to deal with TROVE implementation.

It should be noted that the eigenfunction methods for many-particle harmonic oscillator wavefunctions was also explored by \citet{94NoKaxx}.

%% m = 2

\subsection{Reduction of the rotational rigid rotor basis functions}
\label{s:rotation}

TROVE uses the rigid rotor wavefunctions (Wigner $D$-functions) as the rotational basis set.
In principle, for most of the groups (such as  \Cv{n}, \Ch{n}, \Dh{n}, or \Dd{n}) the symmetry properties of the rigid rotor wavefunctions $\ket{J,k,m}$ are trivial and can be reconstructed based on the $k$ value only \cite{98BuJexx.method}. This is possible because all symmetry operations from these groups can be associated with some equivalent rotations about the body-fixed  axes $x$, $y$, and $z$  only (see, for example, the discussion in Ref.~\citenum{99BuJexx.CH4}). Furthermore, symmetrized combinations of the rigid-rotor wavefunctions are trivial and can be given by the so-called Wang wavefunctions. For example, TROVE uses the following symmetrization scheme\cite{05YuCaJe.NH3}:
\begin{eqnarray}
\label{e:JK=0}
  \ket{J,0,\tau_{\rm rot}} &=& \ket{J,k,m}, \\
\label{e:JK}
  \ket{J,K,\tau_{\rm rot}} &=& \frac{i^{\sigma}}{\sqrt{2}} \left[ \ket{J,K,m} + (-1)^{J+K+\tau_{\rm rot}} \ket{J,-K,m} \right],
\end{eqnarray}
where $K=|k|$, $\tau_{\rm rot}$ is the value associated with the parity of $\ket{J,K,\tau_{\rm rot}}$, $\sigma  = K$ mod 3 for  $\tau_{\rm rot}=1$,  $\sigma = 0$ for $\tau_{\rm rot} = 0$  \cite{05YuCaJe.NH3}, and $m$ is omitted on the left-hand side for simplicity's sake. The symmetry properties of $\ket{J,K,\tau_{\rm rot}}$ can be derived from the properties of $\ket{J,k,m}$ under the associated rotations \cite{98BuJexx.method} and depend on $J,K$ and $\tau_{\rm rot}$ only. Therefore a more sophisticated  symmetrization  approach like the one presented above is not required in such cases. As an example,  Table~\ref{t:rot-C3v} lists the symmetries of $\ket{J,K,\tau_{\rm rot}}$  for a rigid XY\3-type molecule (\Cv{3}(M)) described above.

\begin{table}
\begin{center}
\caption{\label{t:rot-C3v} \Cv{3}(M) symmetries of the rigid-rotor wavefunctions $\ket{J,K,\tau_{\rm rot}}$ ($K\ge 0 $) for the case of a rigid XY\3\ molecule.  $K=0$ is the special case with  $\tau_{\rm rot} = 0$ (even $J$) and $\tau_{\rm rot} = 1$ (odd $J$).}
\begin{tabular}{ccc}
\hline
$\Gamma$ & $K$ & $\tau_{\rm rot}$ \\
\hline
$A_1        $&$ 3n $& 0 \\
$A_2        $&$ 3n $& 1 \\
$E_a        $&$ 3n \pm 1 $& 0\\
$E_b        $&$ 3n \pm 1 $& 1\\
\hline
\end{tabular}
\end{center}
\end{table}

However, some symmetry groups contain operations with equivalent rotations about other axes than $x$, $y$, and $z$, such as \Td\ and \Oh. Consider a rigid XY\4\ molecule spanning the \Td(M) symmetry group.  The permutation (124) is associated with the equivalent rotation $R_3(1,1,1)$, which is a $2 \pi / 3$ right-hand rotation about an axis from the origin to the point $(x, y, z) =(1, 1, 1)$ \cite{99BuJexx.CH4}. In this case the symmetrized basis can only be formed from a linear combination of $\ket{J,k,m}$ spanning a range of $k$ values, as was also shown by \citet{11AlLeCa.CH4}.  This is where we use the TROVE symmetrization approach to build symmetry adapted rotational basis functions $\ket{J,\Gamma}$ (see also Refs.~\citenum{14YuTeBa.CH4,14YuTexx.CH4}, where this approach was applied for $J$ up to 45). The formulation of the symmetrization scheme is given in the Appendix.

Once the ${\bf D}^{\rm Wang}[R]$ matrices are known, the numerically-adapted reduction scheme described above is used to build the symmetrized representation for any $J$. The rotational quantum number $K$ cannot be  used for classification of these symmetrized rigid-rotor combinations anymore. Instead we label them as $\ket{J,\Gamma,n}$, where $n$ is a counting index.

%\begin{eqnarray}
%% \nonumber to remove numbering (before each equation)
%  \ket{J,i,\Gamma} &=& C_{k}^{J,i,\Gamma} \ket{J,k,m}  \\
%  \ket{J,i,\Gamma} &=& \sum_{k=-J}^{J} C_{k}^{J,i,\Gamma} \ket{J,k,m}  \\
%\end{eqnarray}

\subsection{Constructing (ro-)vibrational basis sets}
\label{s:ro-vib}

Following the subspace-based approach introduced for symmetrization of the vibrational part, the rotational modes are also treated as part of an independent, rotational subspace, which is referred in TROVE to as subspace~0. The symmetry adapted ro-vibrational basis set is then constructed as a direct product of the symmetrized components from different subspaces as  $ \Psi_{\lambda_0}^{(0),\Gamma_0} \otimes
\Psi_{\lambda_1}^{(1),\Gamma_1} \otimes \Psi_{\lambda_2}^{(2),\Gamma_2} \cdots \otimes
\Psi_{\lambda_L}^{(L),\Gamma_L}$, where $L$ is the number of vibrational subspaces. The product of irreducible representations must be further reduced, which is much easier when each component is transformed as one of the irreps of the group with standard transformation properties. In this case the same projection operator symmetrization technique is used without further sampling of the symmetric properties of the corresponding components.

%Different rotational angular momenta $J$ thus correspond to different ro-vibrational basis sets.

An efficient alternative to the vibrational basis set as a direct product of subspaces is the $J=0$ contraction scheme \cite{09YuBaYa}. According to this scheme the eigenfunctions of the vibrational ($J=0$) problem are used as contracted vibrational basis functions
for $J>0$. The $J=0$ eigenfunctions represent an even more compact basis set and can be further contracted (referred to as the $J=0$ contraction). The symmetry adapted ro-vibrational basis set is then constructed exactly as described above (using the same numerical symmetrization subroutines) as a direct product of $\Phi_{\lambda_0}^{(0),\Gamma_0} \otimes \Phi_{\lambda_1}^{(1),\Gamma_1}$, where the subspace-index $i$ in $\Phi_{\lambda_i}^{(i),\Gamma_{i}}$ runs over 0 and 1 only, and the $J=0$ basis functions are combined into subspace~1.

%\section{Appendix}

\section{Conclusion}
\label{s:concl}

A new method for constructing symmetry adapted basis sets for ro-vibrational calculations has been presented. The method is a variation of the matrix (or eigenfunction) approaches and is based on solving eigenfunction problems for a set of reduced Hamiltonian operators without resorting to rigorous group-theoretical algebra.
%Our method can be seen as a variation of the CSCO eigenfunction symmetrization method \citep{85ChGaMa.method} method based on the so-called complete set of commuting operators.
% instead of employing symmetrization operators constructed solely from the symmetry transformations.
The advantage of using reduced Hamiltonians in the matrix symmetrization is that it also improves the properties of the basis sets by making them more compact and adjusted to the physics of the problem, thus allowing for efficient contraction. However, it lacks the automatic classification of the basis functions by the irreps, which is a useful feature of the CSCO-based eigenfunction approach by \citet{85ChGaMa.method}. To make up for this, the TROVE symmetrization procedure has to be complemented by a sampling technique accompanied by a projection-based reduction.

Our symmetrization approach has been implemented in the TROVE program suite and has been extensively used for a variety of tri-, tetra-, and penta-atomics covering the \Cs(M), \Cv{2}(M), \Cv{3}(M), \Dh{2}(M), \Dh{3}(M), and \Td(M) groups. TROVE symmetrization is general in that it can be applied to any molecule with arbitrary selection of coordinates provided the symmetry properties of the latter are known. We are now implementing a general numerical technique for \Ch{n}(M), \Cv{n}(M), and \Dh{n}(M) representations, where $n$ is an arbitrarily large integer value. Although TROVE symmetrization was developed and used for building ro-vibrational basis sets, we believe it can be useful for many other applications in physics and chemistry. The symmetrization subroutines (Fortran 95) are written to be as general as possible and in principle can be interfaced with other variational codes, if there will be interest from the community. The illustration  of the symmetrization approach applied to the Harmonic oscillator wavefunctions (see Appendix) is an example where using TROVE is not necessary and thus could be a good place to start.

%\Que{The code written by the authors for constructing symmetry adapted bases were extremly
%useful if it could be interfaced with any other variational ro-vibrational code (GENIUSH, for
%nstance) easily. It is not clear from the ms if this is the case.} \Ans{Comment?}

%Standard (\red{\bf what about non standard but same for all states (same as first encountered %degenerate state)?}: we should make a note about this in conclusion)

\section*{Acknowledgements}

This work was supported by
%the UK Science and Technology Research Council (STFC) No. ST/M001334/1 and
the ERC Advanced Investigator Project 267219. We also thank the support of the COST action MOLIM No. CM1405,  UCL for use of the Legion High Performance Computer and DiRAC@Darwin HPC cluster. DiRAC is the UK HPC facility for particle physics, astrophysics, and cosmology and is supported by STFC and BIS. SNY and RIO thank Per Jensen for very helpful discussions and inspiration. The work of RIO was supported by RFBR No. 15-02-07473, 15-02-07887, and   16-32-00668.
We also thank Alec Owens for proofreading the manuscript. His valuable comments and suggestions led to a significant improvement of this work and are greatly appreciated.

\clearpage

\appendix

\section{Symmetrized 2D Harmonic oscillator basis sets}
\label{a:harmonic}

Here we illustrate how to build a \Dh{3}(M) symmetrized vibrational basis set for ammonia-type molecules to represent the two asymmetric bending modes $q_4$ and $q_5$ from subspace~2 (see Eqs.(\ref{e:q4:NH3},\ref{e:q5:NH3})) for the example from Section~\ref{s:XY3:degen}. The basis set will be formed from the products of the degenerate Harmonic oscillator basis set functions as given by:
$$
 \phi_{n_4,n_5}^{\rm HO}(q_4,q_5) = C_{n_4,n_5} H_{n_4}(q_4) e^{-\alpha q_4^2} H_{n_5}(q_5) e^{-\alpha q_5^2}.
$$
%$$
%\phi_{n_i}^{\rm HO}(q_i) = C_{n_i} H_{n_i}(q_i) \, e^{-\alpha q_i^2}
%$$
where $H_{n}$ is a Hermite polynomial,  $\alpha$ is a parameter (the same for all degenerate components), and $C_{n_4,n_5}$ is a normalization constant.
% defined as
%\red{xxxxxx - CHECK!}
%$$
%\alpha = \omega^2 / g_0. ????
%$$
%For the ammonia example presented above, we now apply the Harmonic oscillators as basis functions %for the two bending modes $q_5$ and $q_6$ (subspace~2) and build a symmetry adapted representation %from the product basis defined by
%$$
% \phi_{n_4,n_5}^{\rm HO}(q_4,q_5) = H_{n_4}(q_4) e^{-\alpha q_4^2} H_{n_5}(q_5) e^{-\alpha q_5^2}.
%$$
%where $\alpha$ is a parameter.
These functions represent solutions of the 2D degenerate Harmonic oscillator
$$
  H^{\rm HO} \phi_{n_4,n_5}^{\rm HO}(q_4,q_5) = \tilde\omega (n_4+n_5+1) \phi_{n_4,n_5}^{\rm HO}(q_4,q_5),
$$
and can be combined to express a solution of the 2D isotropic Harmonic oscillator (IHO):
$$
\Psi_{N,l}^{\rm IHO} = F_{N,l}(\rho) e^{i l \phi},
$$
where \cite{98BuJexx.method}
$$
\rho = \sqrt{q_4^2+q_5^2}, \;\; \phi = \arctan \frac{q_{5}}{q_4},
$$
$$
N = n_4 + n_5,  \;\; l  = N, N-2, ... -N+2, -N.
$$
Here $\Psi_{N,l}^{\rm IHO}$ is an eigenfunction of the corresponding 2D IHO problem:
$$
H^{\rm HO} \Psi_{N,l}^{\rm IHO} =  \tilde{\omega} (N+1) \Psi_{N,l}^{\rm IHO},
$$
which transforms as $A_1/A_2$ (for $l=0,3,6,\ldots$) and $E$ (otherwise).
That is, there exists a linear transformation that connects $\Psi_{N,l}$ and $\phi_{n_4,n_5}(q_4,q_5)$ subject to $N=n_4+n_5$.

In order to find such a transformation and thus build the symmetry adapted functions $\Psi_{N,l}$, we apply the TROVE numerical symmetrization procedure. For example, for a given polyad number  $N=3$, we need to combine the following four products $\phi_{n_4}(q_4)\phi_{n_5}(q_5)$ satisfying  $n_4+n_5=3$:
\begin{equation}\label{eq:4:harm}
 \phi_{3}\phi_0, \quad \phi_{2}\phi_{1}, \quad  \phi_{1}\phi_{2}, \quad  \phi_{0}\phi_{3}.
\end{equation}
These four wavefunctions are degenerate and share the same Harmonic oscillator energy \cite{98BuJexx.method}
$$
\tilde{E}_{n_4,n_5}^{\rm HO} = \tilde{\omega} \, (n_4+n_5+1),
$$
with $\tilde{\omega} = 1679.380$~\cm\ and $\alpha =0.2241$~rad$^{-2}$.
We use a sampling grid of 40 geometries ranging between $-1.0\leq q_4,q_5 \leq 1.0$~radians to probe the values of the  wavefunctions and their symmetric replicas and to obtain the six $4\times 4$ transformation matrices ${\bf D}[R]$ for each operation $R$ in \Dh{3}(M).
%for definition of the symmetry operations $R$ in \Dh{3}(M), their characters and the transformational properties of $q_i$).

Applying the group operations $E$, $(123)$, $(23)$, $E^*$, $(132)^*$, and $(23)^*$ to the four selected degenerate wavefunctions $\phi_1 = \ket{0,3}, \phi_2=\ket{1,2}, \phi_3=\ket{3,0}, \phi_4=\ket{2,1}$, the following matrices ${\bf D}[R]$ are obtained:
$$
{\bf D}[E] =  {\bf D}[E^*] =  \left(
\begin{array}{cccc}
1	&	0	&	0	&	0	\\
0	&	1	&	0	&	0	\\
0	&	0	&	1	&	0	\\
0	&	0	&	0	&	1
\end{array}
\right),
\quad
{\bf D}[(23)] = {\bf D}[(23)^*] \left(
\begin{array}{cccc}
-1	&	0	&	0	&	0	\\
0	&	1	&	0	&	0	\\
0	&	0	&	-1	&	0	\\
0	&	0	&	0	&	1
\end{array}
\right),
$$
$$
{\bf D}[(132)] = {\bf D}[(132)^*] =  \left(
\begin{array}{cccc}
-\frac{1}{8}	&	-\frac{3}{8}	&	-\frac{3\sqrt{3}}{8}	&	-\frac{3\sqrt{3}}{8}	\\
\frac{3}{8}	&	\frac{5}{8}	&	\frac{\sqrt{3}}{4}	&	-\frac{3\sqrt{3}}{8}	\\
-\frac{3\sqrt{3}}{8}	&	-\frac{\sqrt{3}}{4}	&	\frac{5}{8}	&	-\frac{3}{8}	\\
\frac{3\sqrt{3}}{8}	&	-\frac{3\sqrt{3}}{8}	&	\frac{3}{8}	&	-\frac{1}{8}	\\
\end{array}
\right),
$$
where the matrix elements are given to $10^{-15}$.
With the help of Eq.~(\ref{e:charcters:relation}) we obtain the reduction coefficients $a_i$  = $1, 1, 1, 0, 0, 0$ (within $10^{-15}$) ($i$ = $A_1\p$, $A_2\p$, $E\p$, $A_1\pp$, $A_2\pp$, $E\pp$), i.e. only  $A_1\p$, $A_2\p$, $E\p$ combinations can be formed. For the 1D representations
$A_1$  and $A_2$ the projection operators obtained using Eq.~\eqref{e:projection} are given by
\begin{eqnarray}
\label{e:Harm:A1}
P^{A_1} &=&  \left(
\begin{array}{cccc}
0	&	0	&	0	&	0	\\
0	&	\frac{3}{4}	&	0	&	-\frac{\sqrt{3}}{4}	\\
0	&	0	&	0	&	0	\\
0	&	-\frac{\sqrt{3}}{4}	&	0	&	\frac{1}{4}
\end{array}
\right), \\
\label{e:Harm:A2}
P^{A_2} &=&  \left(
\begin{array}{cccc}
\frac{1}{4}	&	0	&	-\frac{\sqrt{3}}{4}	&	0	\\
0	&	0	&	0	&	0	\\
-\frac{\sqrt{3}}{4}	&	0	&	\frac{3}{4}	&	0	\\
0	&	0	&	0	&	0
\end{array}
\right).
\end{eqnarray}
These matrices contain a total of eight vectors that we can choose from to build the irreducible combinations of $\phi_1$, $\phi_2$, $\phi_3$ and $\phi_4$, four of which are trivial and only two pairs are linearly independent. Choosing the second column vector from the $P^{A_1}$ matrix, after normalization we obtain
$$
\Psi_{1}^{A_1}(q_4,q_5) =  \frac{\sqrt{3}}{2} \ket{1}\ket{2} - \frac{1}{2} \ket{3}\ket{0},
$$
where the index 1 indicates the counting number of this state. The only  non-trivial and linearly independent choice is given by (after normalization):
$$
\Psi_{2}^{A_2}(q_4,q_5) = \frac{1}{2} \ket{0}\ket{3} - \frac{\sqrt{3}}{2} \ket{2}\ket{1}.
$$

The projection operator for the $E$-symmetry component leads to the following  matrices:
\begin{eqnarray}
\label{e:Harm:E11}
P^{E}_{11} &=&  \left(
\begin{array}{cccc}
0	&	0	&	0	&	0	\\
0	&	\frac{1}{4}	&	0	&	\frac{\sqrt{3}}{4}	\\
0	&	0	&	0	&	0	\\
0	&	\frac{\sqrt{3}}{4}	&	0	&	\frac{3}{4}
\end{array}
\right), \\
\label{e:Harm:E22}
P^{E}_{22} &=&  \left(
\begin{array}{cccc}
\frac{3}{4}	&	0	&	\frac{\sqrt{3}}{4}	&	0	\\
0	&	0	&	0	&	0	\\
\frac{\sqrt{3}}{4}	&	0	&	\frac{1}{4}	&	0	\\
0	&	0	&	0	&	0
\end{array}
\right).
\end{eqnarray}
From this space of eight possible vectors we select the following trial vector, which is linear independent from  $\Psi_{1}^{A_1}$ and $\Psi_{2}^{A_2}$:
$$
\phi^{\rm trial} = (0,\frac{3}{4},0,\frac{\sqrt{3}}{4})^{T}.
$$
This vector is then orthogonalized to $\Psi_{1}^{A_1}$ and $\Psi_{2}^{A_2}$ using the Grand-Schmidt procedure to get
$$
\Psi_{3,1}^{E_a}(q_4,q_5) =  \frac{1}{2} \ket{1}\ket{2} + \frac{\sqrt{3}}{2} \ket{3}\ket{0}.
$$
The second component of the latter  vector, $E_b$, is obtained by applying the corresponding transfer operator in Eq.~\eqref{e:transfer} to $\Psi_{3,1}^{E_a}$, which after the Grand-Schmidt orthogonalization reads:
$$
 \Psi_{3,2}^{E_b}(q_4,q_5) =   -\frac{\sqrt{3}}{2} \ket{0}\ket{3} - \frac{1}{2} \ket{2}\ket{1}.
$$
In all these equations above, the coefficients $1/2$, $3/4$, $\sqrt{3}/2$ and $\sqrt{3}/4$  are obtained numerically and they coincide with the numerical values to within $10^{-14}$.

This completes the example of the reduction of degenerate wavefunctions in the basis of 2D isotropic Harmonic oscillator functions. It is also a good illustration of how our method is applied to degeneracies of arbitrary order and dimensions.
The only limitation is the memory and time required for sampling wavefunctions and inverting the transformation matrices via Eq.~(\ref{e:represent-2}). For example in Ref.~\citenum{13YuTeBa.CH4} the highest degeneracy order used was 120 together with the 2D and 3D symmetries $E$, $F_1$ and $F_2$.

It should be noted that as an alternative to the orthogonalization with Grand-Schmidt, one can impose the orthogonality conditions on the elements of matrix ${\bf D}[R]$ during the solution of Eq.~(\ref{e:represent-combine}), for example by utilizing the exponential ansatz:
\begin{eqnarray}
{\bf D}[R] = \exp(-{\boldsymbol\kappa}[R]),~~~{\boldsymbol\kappa}^T=-{\boldsymbol\kappa},
\end{eqnarray}
with only one independent element $\kappa_{12}$ in case of a doubly-degenerate irrep, three independent elements $\kappa_{12}$, $\kappa_{13}$, and $\kappa_{23}$ in case of a triply-degenerate irrep, etc. to be determined.
Using this representation, the system of equations in Eq.~(\ref{e:represent-combine}) becomes nonlinear and can be easily solved using the iterative approach described in Ref.~\citenum{15YaYuxx.method} for solution of the Eckart equations. We are planning to explore this approach in the future.

\section{Symmetrization of the rotational rigid rotor wavefunctions}
\label{a:rotation}

The symmetry transformation properties (i.e. transformation matrices ${\bf D}[R]$) of the rigid-rotor wavefunctions required for our symmetrization scheme can be obtained directly from the Wigner $D$-functions, associated with the corresponding Euler angles of the particular equivalent rotation $R(\alpha,\beta,\gamma)$. These are given by (see also \cite{11AlLeCa.CH4}):
$$
 R(\alpha,\beta,\gamma) \ket{J,k,m} = \sum_{k\p=-J}^{J} {D_{k\p,k}^{(J)*}}(\alpha,\beta,\gamma) \ket{J,k\p,m}.
$$
Thus the sampling procedure is not required for rotational basis functions.
For example, the Euler angles for all equivalent rotations of a rigid XY\4\ molecule (\Td(M)) are given in Table~4 of Ref.~\citenum{11AlLeCa.CH4}.
The transformation properties of the Wang functions $\ket{J,K,\tau_{\rm rot}}$ in Eqs.~(\ref{e:JK=0},\ref{e:JK}) can then be deduced using the unitary transformation
%\red{CHECK CONJUGATION HERE!}
$$
{\bf D}^{\rm Wang}[R] = {\bf U}^{+} {{\bf D}^{(J)}}^{*}(\alpha,\beta,\gamma) \bf{U},
$$
where the $(2J+1)\times (2J+1)$ matrix $U_{i,j}$ is given by
$$
U_{1,1} = \left\{
\begin{array}{cc}
         1,& {\rm even} \; J,  \\
  -(i)^{J},& {\rm odd} \; J
\end{array}
\right.
$$
and
\begin{eqnarray}
U_{n,n} & = & \frac{1}{\sqrt{2}}, \\
U_{n,n+1} & = & -i  \frac{(-1)^{\sigma}}{\sqrt{2}}, \\
U_{n+1,n} & = & \frac{(-1)^{J+K}}{\sqrt{2}}, \\
U_{n+1,n+1} & = & i  \frac{(-1)^{J+K+\sigma}}{\sqrt{2}}, \\
U_{n,n'} & = & 0, \quad {\rm for} \quad |n-n'| >1,
\end{eqnarray}
where $n= 2 K$, $K=1\ldots J$, $\sigma  = K$ mod 3 for  $\tau_{\rm rot}=1$,  and $\sigma = 0$ for $\tau_{\rm rot} = 0$.
Once the transformation matrices are known, the standard projection technique described above is applied to obtain the symmetry adapted rigid-rotor combinations used as rotational basis functions \citep{14YuTexx.CH4}.

%\begin{eqnarray}
%% \nonumber to remove numbering (before each equation)
%  \ket{J,i,\Gamma} &=& C_{k}^{J,i,\Gamma} \ket{J,k,m}  \\
%  \ket{J,i,\Gamma} &=& \sum_{k=-J}^{J} C_{k}^{J,i,\Gamma} \ket{J,k,m}  \\
%\end{eqnarray}

%\bibliography{journals_phys,methods,CH4,CH4_,NH3,NH3_,PH3,PH3_,SiH4,H2O2,H2O2_,H2S,H2S_,H3+,H2CO,H2CO_,15NH3,linelists,CH3,SO3,SO3_,SbH3,OH3+,CH3Cl,CH3Cl_,HNO3_,methods_,programs,programs_,symmetry}

\providecommand{\latin}[1]{#1}
\makeatletter
\providecommand{\doi}
  {\begingroup\let\do\@makeother\dospecials
  \catcode`\{=1 \catcode`\}=2\doi@aux}
\providecommand{\doi@aux}[1]{\endgroup\texttt{#1}}
\makeatother
\providecommand*\mcitethebibliography{\thebibliography}
\csname @ifundefined\endcsname{endmcitethebibliography}
  {\let\endmcitethebibliography\endthebibliography}{}

\end{document}